\documentclass[10pt,pra,aps,twocolumn,amsmath,amssymb,showpacs,showkeys,floatfix]{revtex4-2}
\usepackage{graphicx}
\usepackage{amsfonts}
\usepackage{epsfig}
\usepackage{tikz}
\usetikzlibrary{quantikz2}
\usepackage{dcolumn}
\usepackage{amsthm}
\usepackage{color}
\usepackage{subcaption}
\usepackage{braket}
\usepackage{physics}
\usepackage{latexsym}
\usepackage{amscd}
\usepackage[mathscr]{eucal}

\usepackage{algpseudocode}
\usepackage{algorithm}

\usepackage{mathtools}
\usepackage{placeins}
\DeclarePairedDelimiter\ceil{\lceil}{\rceil}
\DeclarePairedDelimiter\floor{\lfloor}{\rfloor}
\begin{document}
\newcommand{\newc}{\newcommand}

%\newc{\beq}{\begin{equation}}
%\newc{\eeq}{\begin{equation}}
%\newc{\ovl}{\overline}
%\newc{\bc}{\begin{center}}
%\newc{\ec}{\end{center}}
%\newc{\tr}{\mbox{tr}}
\newc{\pd}{\partial}
\newc{\dqv}{\delta\vec{q}}
\newc{\dpv}{\delta\vec{p}}
 \newc{\f}{\frac}

\title{Amplitude Amplification and Estimation using a Floquet system}
\author{Keshav. V}
\email{keshav.v@students.iiserpune.ac.in}
\author{M. S. Santhanam}
\email{santh@iiserpune.ac.in}
\affiliation{Indian Institute for Science Education and Research Pune\\ Dr. Homi Bhabha Road, Pune 411 008, India.}

\date{\today}

\noindent\begin{abstract}
The quantum kicked rotor (QKR) is a fundamental model of time-dependent quantum chaos and the physics of Anderson localization. It is one of the most well-studied Floquet systems. In this work, it is shown that QKR can be used to implement a quantum algorithm to perform unstructured search; namely Amplitude Amplification, a generalization of Grover's search algorithm. Further, the QKR is employed for amplitude estimation when the amplitude of the marked states is unknown. It is also shown that the characteristic property of dynamical localization of the QKR can be exploited to enhance the performance of the amplitude amplification algorithm by reducing its average runtime. The sensitivity of the success probability of unstructured search to detuning from resonance and the effects of noisy kick strengths are analyzed and the robustness of the QKR based algorithm is demonstrated. The experimental feasibility of every component of the algorithm is discussed.

\end{abstract}
%\pacs{05.45.Mt, 03.65.Ud, 03.67.-a}
\maketitle
\section{Introduction}
\label{sec:Introduction}
Amplitude Amplification \cite{Brassard_2002, Hoyer_2000} is a versatile and popular technique that powers quantum algorithms. It is a deterministic scheme to systematically and selectively amplify the amplitudes of known or unknown states, starting from an initial superposition state. If we are searching for $M$ items in an unstructured database with $N$ items, then amplitudes corresponding to the items of interest can be amplified to nearly unity in $O(\sqrt{N/M})$ calls to an "oracle" which encodes these items, as opposed to the best classical algorithm, which requires $O(N/M)$ Oracle calls. This technique is based on a generalization of Grover's algorithm for unstructured database search \cite{grover1996fast}, and is prototypical example that delivers a quadratic speedup over the best known classical algorithm. Grover's original algorithm \cite{grover1996fast} was a specific scheme to enhance the amplitude of the ``marked'' state, involving repeated applications of a multi-qubit Hadamard gates (often called the ``inversion about the mean'') and a Grover oracle. It was also subsequently realized that the Hadamard gates could be substituted with almost any unitary operator to amplify amplitudes provided that there is a scheme for implementing the Grover oracle and time-reversed dynamics \cite{PhysRevLett.80.4329}. This is an immensely useful, though it requires an additional amplitude estimation procedure to estimate the number of iterations needed to reach success probability of $O(1)$ \cite{Brassard_2002}.

Over the years, Grover's algorithm has been experimentally implemented in various test-beds. The first implementation utilized a nuclear magnetic resonance system with 4-qubits \cite{ChuGerKub1998}, while a later attempt significantly improved the effciency of readout \cite{DAS20038}. Other significant implementations include a 2-qubit experiment based on trapped atomic ions \cite{BriHalLee2005, Figgatt_2017}, using a system of superconducting coupled transmon qubits \cite{DicChoGam2009}, and through a one-way quantum computing paradigm using polarization states of photons \cite{WalResRud2005}, and more recently a variant of Grover was realized with linear optical system \cite{WanChiYu2021, HeZhaLv2023}. Further, a classical implementation in Fourier optics setup shows that quantum speedup can be realized in a far-from-quantum regime \cite{BhaHeuSpr2002}. In many of these platforms, there is usually a compromise between scalability and control. Experimentally controlling larger number of qubits is a challenge, and hence scalability with the system size is dependent on the ability to control the collection of qubits. In this article, we propose a scheme to efficiently implement amplitude amplification and estimation using the quantum kicked rotor (QKR) and argue that it is well suited to overcome some of these challenges.

Over the two decades, a significant interest has been focused on periodically driven quantum systems (also known as Floquet systems) \cite{OkaSot2019}. They are known to exhibit many counter-intuitive effects \cite{Bukov_2015} such as dynamical stabilization \cite{Tiwari_2024}, disorder-free localization \cite{CasChi-qcbook}, and discrete time crystalline order \cite{PhysRevX.12.031037}. Thus, Floquet engineering is a tool to engineer novel interactions and effects. Amplitude amplification involves repeated application of a ``Grover iteration'' operator, making it a Floquet system. Hence, a natural question is whether another Floquet system can be engineered to implement the amplification algorithm. We demonstrate that this is indeed possible using the QKR and some of its variants.

The one-dimensional kicked rotor is a paradigmatic model of Hamiltonian chaos and has been extensively investigated in the last five decades \cite{SANTHANAM20221,Izrailev-1990,Chiri-LesHou,CasChi-qcbook}. It can be thought of as a free particle on a ring that is periodically kicked by an external sinusoidal field. For sufficiently strong kicks, the Classical kicked rotor displays chaos and the particle absorbs unbounded energy. In contrast, in the corresponding quantum regime, the energy absorption is arrested by quantum interferences resulting in the localization of the wave packet in momentum space. This emergent phenomenon, termed dynamical localization, is the momentum space analogue of Anderson localization, and has been experimentally observed in laser-kicked cold atomic clouds and in dilute Bose-Einstein condensates \cite{MooRobBha1994,MooRobBha1995,SarPauVis2017,SANTHANAM20221}. For the present purposes in this work, parameters must be chosen such that dynamical localization is avoided. This is achieved through quantum resonance, {\it i.e.}, by tuning the kick period $T$ to its resonant value $T_r$ such that the free evolution between consecutive kicks becomes an identity operation. For this special case of quantum resonance with $T=T_r$, as proposed and experimentally demonstrated in Refs. \cite{Dadras_2018,DadGreGro2019,DelPetWim2020,DelGroPet2020}, the dynamics induced by QKR with a cosine kick potential is identical to that of a walker in a tight-binding lattice executing a continuous-time quantum walk. In this scenario, the mean energy does not saturate, but instead displays a ballistic growth : $\langle E \rangle \sim t^2$.

The connection between QKR {\sl at resonance} and continuous time quantum walk has been exploited to demonstrate a search protocol on momentum lattice using Bose-Einstein condensates kicked by an optical lattice \cite{DelGroPet2020}. This scheme suffers from the following drawbacks -- the final probability (before measurement) is quite small $\sim 0.1$, and translates to poor fidelity of the search result. Secondly, this scheme involves only one oracle implementation and cannot be mapped to the known optimal quantum algorithm (such as due to Grover) for the same procedures. Since the quadratic speedup in Grover's search is optimal \cite{Boyer_1998}, the scheme reported in Ref. \cite{DelGroPet2020} would not be able to yield a good fidelity without a number of oracle calls quadratic in the number of elements involved in the search. 

In this work, we propose a search scheme based on QKR at resonance that overcomes these issues and, importantly, can be mapped to amplitude amplification and Kitaev's phase estimation protocols. As demonstrated in a later section, this leads to a high fidelity of the search result as expected from a realization of Amplitude Amplification. The procedure has the potential to be more scalable than some other approaches\cite{ChuGerKub1998, DAS20038, BriHalLee2005, Figgatt_2017, DicChoGam2009, WalResRud2005, WanChiYu2021,HeZhaLv2023} since the wavepacket can be made to spread over a large number of momentum lattice sites. We begin by discussing QKR and quantum resonance in Sec. \ref{sec:The kicked rotor and Quantum walks}. In Sec. \ref{sec: Requirements for Amplitude Amplification}, implementation of the QKR-based unitary gates and a procedure for amplitude amplification is outlined. The potential loss of a quadratic speedup and ways to resolve this are discussed in Sec. \ref{sec: Optimizing the inital distribution}, alongside a scheme for amplitude estimation in Sec. \ref{sec: Estimating the required number of iterations}. The effects due to errors on the fidelity of the algorithm is discussed in Sec. \ref{sec: Effects of erroneous detuning from resonance}-\ref{sec: Sensitivity to Noisy kicks}. Finally, in Sec. \ref{sec: Conclusions and outlook}, some potential generalizations to similar kicked systems as well as the experimental protocols are discussed.

\section{Quantum kicked rotor}
\label{sec:The kicked rotor and Quantum walks}
The Hamiltonian of a QKR is given by
\begin{equation}
\widehat{H} = \frac{L^2}{2I} + \hbar\phi ~ V(\theta) ~ \sum_{j=0}^{\infty} \delta(t-jT),
\label{ham1}
\end{equation}
where $\theta$  and $L$ represent, respectively, the angle (position) and angular momentum of the rotor, and $I=\hbar=1$ its moment of inertia. Further, $T$ is the time period of the kicks, $\hbar\phi$ is the kick strength, and $V(\theta)$ is the kick potential.
It describes the dynamics of a periodically kicked particle on a ring implying periodic boundary condition in position alone. In a later section, we employ an additional internal spin-$\frac{1}{2}$ degree of freedom (in the form of two hyperfine states) necessary for achieving phase kickbacks for amplitude estimation. Usually, $V(\theta) = \cos\theta$, but in this work different periodic functions expressed as Fourier series will also be used. 

Since Eq. \ref{ham1} is periodic in time, the dynamics of QKR can be conveniently studied using a Floquet operator
\begin{equation}
    U_{\rm kr} = U_{\rm kick} ~ U_{\rm free} = \exp(-i\phi V(\theta)) ~ \exp(-iL^2T/2\hbar I),
    \label{floq1}
\end{equation}
which evolves an arbitrary initial state for one full time period $T$ starting from just after a kick ending just before the next kick. When $T \ne T_r$, then QKR exhibits dynamical localization in momentum space. Then Eq. \ref{floq1} can be mapped to a Anderson-type model with diagonal disorder and short-range hopping probabilities \cite{PhysRevLett.49.509}. Hence, the dynamical localization in momentum space of QKR is understood to be a counterpart of Anderson localization. Dynamical localization leads to the, exponential profile of eigenstates of $U_{\rm kr}$, a signature of Anderson localization. 
%By now, it is well appreciated that Anderson localization is a generic feature of wave propagation in disordered media \cite{PhysRev.109.1492}. 

\begin{figure}[t]
  \centering
  \includegraphics*[width=\columnwidth]{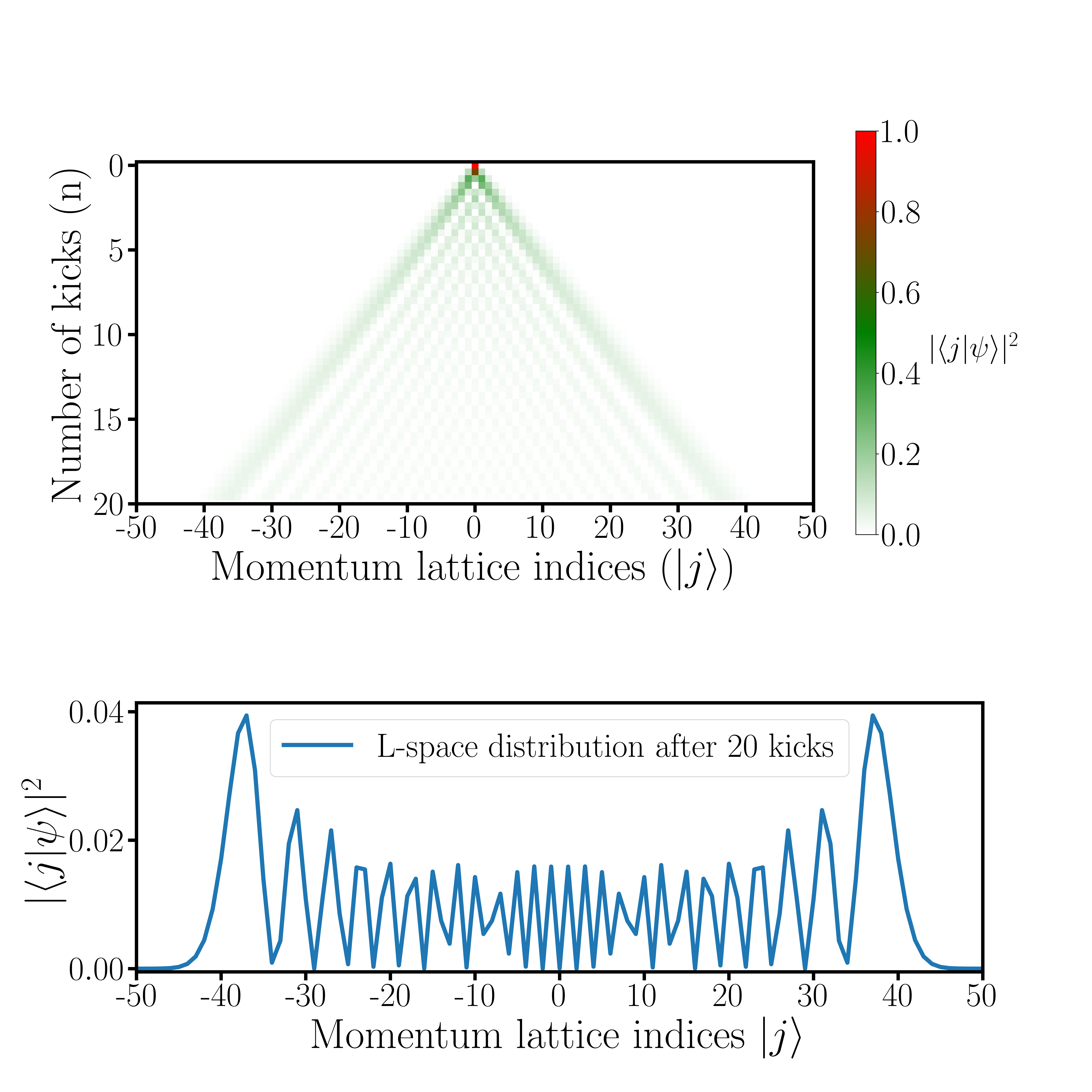}
  \caption{Dynamics of quantum kicked rotor under conditions of quantum resonance with kick strength $\phi$=2. (a) space-time density plot over momentum lattice points, and (b) density profile at $t=20$ over momentum lattice. This clearly resembles that of a continuous-time quantum  walk.}
  \label{fig:QKR-CTQW probability distribution}
\end{figure}

In contrast, if we choose $T=T_r = 4\pi I/\hbar$ (condition for quantum resonance), then the free evolution becomes an identity operator. This is because the spectrum of the angular momentum operator is quantized, owing to the periodic boundary conditions, as $n\hbar$. The temporal evolution generated by the unitary operator is
\begin{equation}
    U = U_{\rm kick} = e^{-\phi V(\theta)}.
    \label{u_qres}
\end{equation}
 It can then be shown that a ballistic growth in the mean energy with time is also observed. As is usually done, in this work, $\hbar=1$ and the kick period $T$ is tuned to satisfy resonance condition. It might be noted that if $V(\theta) = \cos \theta$, then the resonant QKR generates a continuous-time Quantum walk in the momentum space of an infinite path graph with each node being an eigenstate of the angular momentum operator \cite{DelGroPet2020}.

This property is useful in generating an initial state that is approximately a uniform superposition of basis states. Figure \ref{fig:QKR-CTQW probability distribution}(a) shows the space-time plot for a quantum walk generated by the kicked rotor Hamiltonian Eq. \ref{ham1} at resonance. The profile has the familiar conical shape reflecting that the standard deviation of the walk is $\propto n$, the number of kicks and is similar to that for a continuous-time quantum walk (CTQW). As seen in \ref{fig:QKR-CTQW probability distribution}(b), the probability distribution over momentum space at $n=200$ kicks is quite rugged and is consistent with that expected for a CTQW. In principle, it would be possible to implement the amplitude amplification algorithms with almost any kick potential, but a quadratic speedup can be realized only when the amplitudes of initial state are reasonably uniform in magnitude.

\section{QKR for amplitude amplification}
\label{sec: Requirements for Amplitude Amplification}
In this section, we will discuss an algorithm for amplitude amplification using the QKR dynamics at resonance. The first step of the Grover's algorithm creates an initial state uniformly distributed over the computational basis, and it is usually achieved by application of Hadamard operators. In the generalized amplitude amplification algorithm demonstrated here, the following steps will be performed using QKR unitary operators; (a) A superposition state will be created by $U$ in Eq. \ref{u_qres}, (b) States are marked using the Grover oracle and (c) Time-reversed dynamics are executed using $U^{\dag}$. 
These steps can be stated as an algorithm as follows :
\begin{algorithm}[H]
\caption{Amplitude Amplification}
\begin{algorithmic}[1]
\State \textbf{Initialization} Perform N kicks to get a superposition state $\ket{\psi} = e^{-i\phi V(\theta)}\ket{0}$
\While{$i \leq \left\lfloor \frac{\pi}{4 \sin^{-1} \sqrt{a}} \right\rfloor$}
   \State Marking: Apply the Oracle O
   \State Apply $U^\dagger$ by shifting the phase of the kick potentials
   \State Mark the state $\ket{0}$ using the operator $O = 1 - 2\ket{0}\bra{0}$
   \State Apply U using the usual kick potentials
   \State $i = i + 1$
\EndWhile
\State Measure the system in the momentum basis
\end{algorithmic}
\end{algorithm}
In the following subsections, we deal with each one of these requirements. Note that steps 4-6 in this algorithm together constitute the Grover diffusion operation.

\subsection{Creating an optimal initial distribution}
\label{sec: Optimizing the inital distribution}
In principle, algorithm I can be implemented using a cosine kick potential, Eq. \ref{u_qres} with $V(\theta)=\cos\theta$, and any momentum state can be marked. However, in practise, this may not always lead to a quadratic speedup because this potential produces a highly non-uniform probability distribution in momentum space. If the marked site has a low amplitude, the number of iterations required for its amplification, and the average runtime might not have the $\sqrt{N}$ scaling. Furthermore, even if the number of marked sites is known, their identities are not. It is therefore not possible to know the required number of iterations, which depends on the amplitudes of the marked states. For these reasons, the initial distribution must be made as close to uniform as possible. To this end, we present two distinct strategies to smoothen out the distribution and generate a closer-to-uniform initial state.
\begin{figure}[t]
     \centering
     \includegraphics[width=0.85\columnwidth]{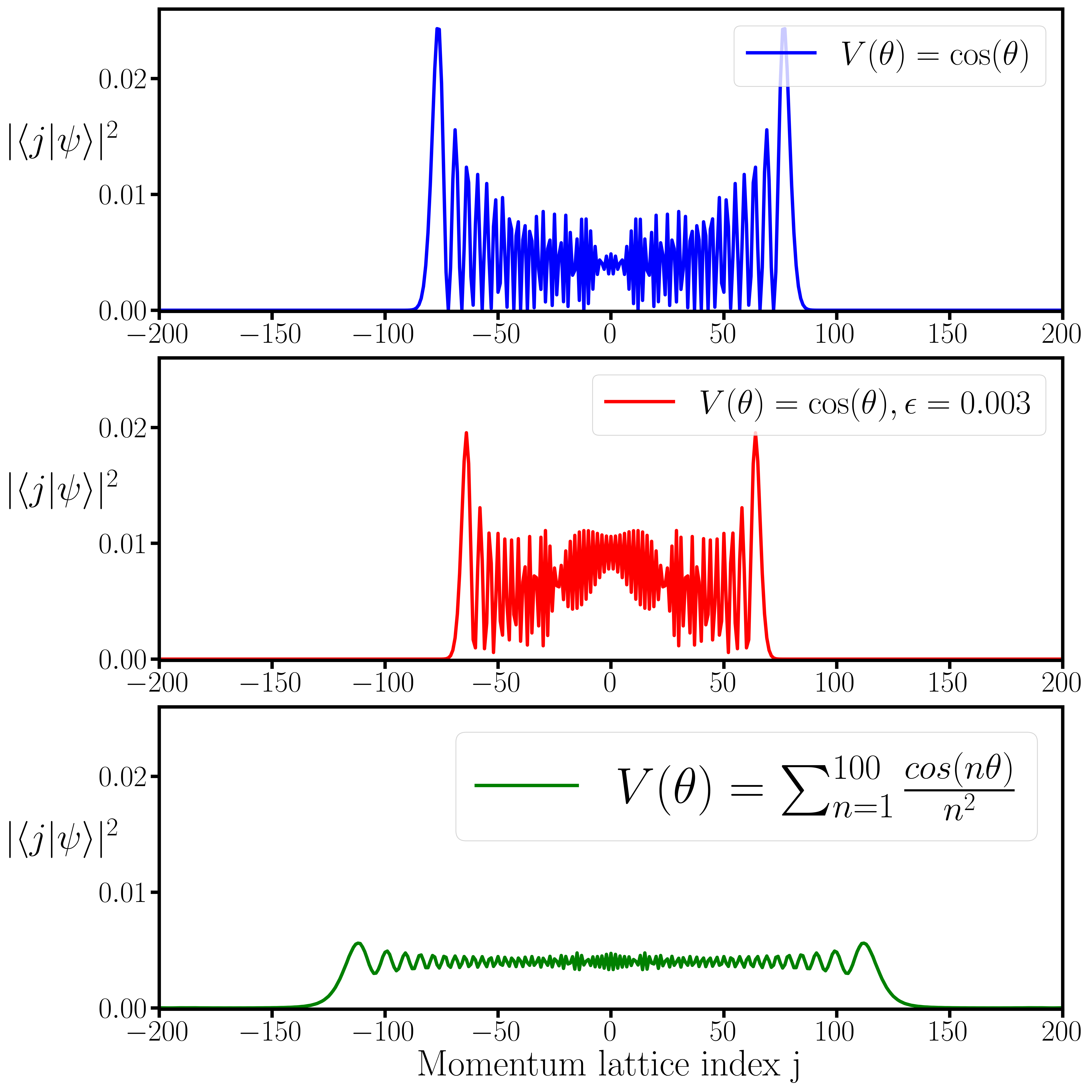}
     \caption{Initial probability distributions in momentum space using resonant QKR after 2 kicks. The initial profile obtained using the modified potential (bottom) is closer a uniform distribution and spreads over momentum basis more than in the other two cases.}
     \label{fig:Density profiles from detuning and M.P compared}
\end{figure}

{\it Modified potential approach:} Numerically, it is found that kick potentials of the form 
\begin{equation}
V(\theta)= \sum_{m=1}^M \frac{\cos m \theta}{m^2}
\label{eq:modpot}
\end{equation}
produce a near-uniform distribution with fluctuations that become insignificant for $M >> 1$. Moreover, marking multiple sites might allow the fluctuations in amplitudes of individual sites about the mean to average out, and this in turn might allow for quadratic speedup to be realized. The momentum-space probability density profile for this scenario with $M=100$ is shown in Fig. \ref{fig:Density profiles from detuning and M.P compared} while the amplitude amplification procedure performed using the modified kick potential is visualized in Fig. \ref{fig:AA using modified potential}.

{\it Approach via controlled detuning:} If detuning from resonance can be precisely controlled, it is possible to engineer an initial momentum-space probability distribution that is more uniform than the one which results from applying resonant kicks alone. The rotor must be allowed to evolve freely in a pre-determined and controlled way in the forward cycle, and the order of the free evolution and the kicks must be reversed in each backward evolution cycle in order to execute time-reversal. This technique was also used previously to achieve Loschmidt cooling by means of time-reversal \cite{PhysRevLett.100.044106}. The operator $U$, representing forward evolution by two non-resonant kicks, is
\begin{equation}
U=e^{-i\phi\cos(\theta)} ~ e^{\frac{-iL^2\epsilon}{2I\hbar}} ~ e^{-i\phi\cos(\theta)}.
\end{equation}
In this, $0 < \epsilon \ll 1$ is the detuning from resonance. The term between the two kicks corresponds to free evolution for time interval $\epsilon$, and prevents the wave-packet from spreading out after the first kick. For $\epsilon \gg 0$, the wave packet localizes in momentum space due to dynamical localization. On the other hand, if $\epsilon$ is kept small, wave packet will spread less than it would have in the resonant case, but the momentum space density profile would be more uniform, as shown in Fig. \ref{fig:Density profiles from detuning and M.P compared}. The amplitude amplification performed by controlling the detuning is visualized in Fig. \ref{fig:AA using detuned rotor}. From an experimental perspective, the controlled detuning is achievable compared to creating a modified potential of the form in Eq. \ref{eq:modpot} (especially with large values of $m$). But it should be noted that even a slight {\it uncontrolled } detuning from quantum resonance could significanlty reduce the fidelity of the search algorithm. This is discussed in Appendix \ref{sec: Effects of erroneous detuning from resonance}

\subsection{Implementing the Grover Oracle}
It is possible to mark sites (momentum states $\ket{p}$) by shifting their phase by $\pi$. This allows us to implement the Grover oracle, a unitary operator that encodes the boolean function $f: \{-N,-N+1,\cdots N-1,N\} \mapsto \{0,1\}$. It takes value 1 for all marked states and 0 otherwise. The corresponding oracle action on momentum eigenstates must be $O\ket{p}=(-1)^{f(p)} \ket{p}$. The marked states $\ket{p}$ are the ones which have $f(p)=1$ and are said to span the good subspace.

\subsection{Implementing the Grover Diffusion operator}
Grover diffusion operator consists of forward and reversed evolution, and the latter can be implemented by simply shifting the phase of the kicking potential $V(\theta) = \cos \theta$ whilst the system is tuned to resonance. Previously, this property was used for interferometry \cite{PhysRevA.86.043604} and to study the effects of quasi-momenta on resonant dynamics \cite{McDowall2009AFT}. For the time-reversed part of each cycle, each term in this series would have to be individually phase-shifted. Alternatively, the sign of the kick strength $\phi$ could be changed to achieve time-reversed evolution. With this, the Grover diffusion cycle is described by 
\begin{equation}
D= e^{-i\phi V(\theta)} \left( 1-2\ket{0}\bra{0} \right) e^{i\phi V(\theta)} = 1-2\ket{\psi}\bra{\psi}.
\label{grover_aa}
\end{equation}
This can be interpreted as a reflection about the vector orthogonal to $\ket{\psi}$ in the subspace spanned by its normalized projections onto the so-called good and bad subspaces. The crucial factor is that the dynamics generated by  $U$ can be easily reversed by changing the sign of $\phi$.
%\subsection{Proof of correctness}

This procedure is identical to the amplitude  amplification technique except for an overall global phase shift $\pi$ applied in each iteration, which does not affect the success probability. We present the proof of correctness along the lines of amplitude amplification \cite{Brassard_2002}. Consider a Hilbert space of orthonormal set of states $\{\ket{n}\}$. The good (marked) and bad (unmarked) subspaces are defined in terms of a Boolean function $\chi(n)$ such that $\chi(n)=1$ indicates marked and $\chi(n)=0$ unmarked subspaces. Let $\ket{\psi}$ represent a state residing in this space, and is obtained from $\ket{0}$ through
\begin{equation}
   \ket{\psi} = e^{-i\phi V(\theta)}\ket{0} = U\ket{0}.
\label{eq:instate}
\end{equation}
This, in turn, can be expressed as
\begin{equation}
    \ket{\psi}= (P_G+P_B)\ket{\psi} =\ket{\psi_1} + \ket{\psi_0}, ~{\rm and} ~ a = \bra{\psi_1}\ket{\psi_1},
\end{equation}
where $P_G$ and $P_B$ are the projectors onto the good and bad subspaces respectively. We assume that only a finite number of states will be marked, which means that the good subspace $\{\ket{g}\}$ is finite dimensional while the bad subspace $\{\ket{b}\}$ is not. The Grover operator, which represents one-iteration of the amplitude amplification process, is defined as
\begin{equation}
  G=U^\dag O_0 U O = (1-2\ket{\psi}\bra{\psi})(1-2P_G),
\end{equation}
in which the operators $O$ and $O_0$ mark the good states and the $\ket{0}$ state respectively. To simplify the notation we introduce $\ket{g}= \frac{\ket{\psi_1}}{\sqrt{a}}$ and $\ket{b}=\frac{\ket{\psi_0}}{\sqrt{(1-a)}}$. We also define $\theta = \sin^{-1}(\sqrt{a})$. A single iteration can be interpreted as a reflection about the projection of $\ket{\psi}$ onto $\{\ket{g}\}$ followed by a reflection about the vector orthogonal to $\ket{\psi}$. Clearly, even after multiple iterations, the dynamics are restricted to the subspace spanned by the vectors $\ket{g},\ket{b}$. We also define $\theta = \sin^{-1}(\sqrt{a})$. The action of a single Grover iteration on the state vector would be the following.
 \begin{equation}
     G\ket{\psi}= U^\dag O_0 U O\ket{\psi} = -\sin(\frac{3\theta}{2})\ket{g} - \cos(\frac{3\theta}{2})\ket{b} 
 \end{equation}
After $r$ iterations, the state would become
\begin{equation}
\ket{\psi^r} = (-1)^r \left( -\sin((2r+1)\theta)\ket{g} -\cos((2r+1)\theta) \ket{b} \right).
\end{equation}
Therefore, we must measure the momentum after 
\begin{equation}
r= \ceil*{\frac{\pi}{4\sin^{-1}(\sqrt{a})} -\frac{1}{2}}
\label{iters}
\end{equation}
iterations of the Grover operator. If the initial state in Eq. \ref{eq:instate} is uniformly distributed in the momentum space, then the probability $a$ for obtaining a marked site will be a good approximation to the fraction of marked sites. In this case, $ r \approx \frac{\pi\sqrt{N}}{4\sqrt{M}}$, indicating that quadratic speedup has been obtained.
\\
\begin{figure}[t]
    \centering
    \begin{subfigure}{\columnwidth}
        \centering
        \includegraphics[width=\columnwidth]{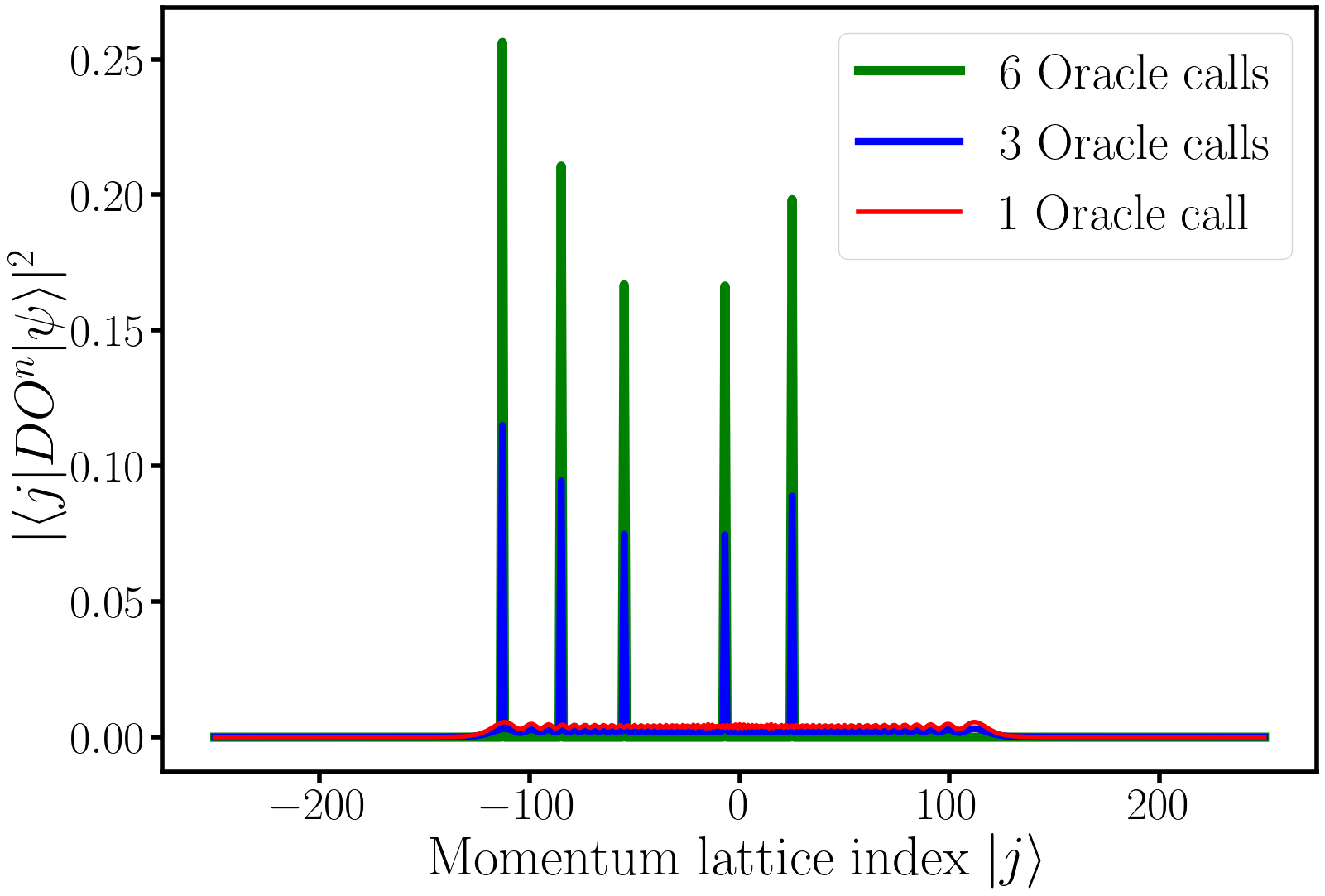}
        \caption{Probability distribution after a different numbers of oracle calls}
    \end{subfigure}
    \hfill
    \begin{subfigure}[b]{1\columnwidth}
        \centering
        \includegraphics[width=1.1\columnwidth]{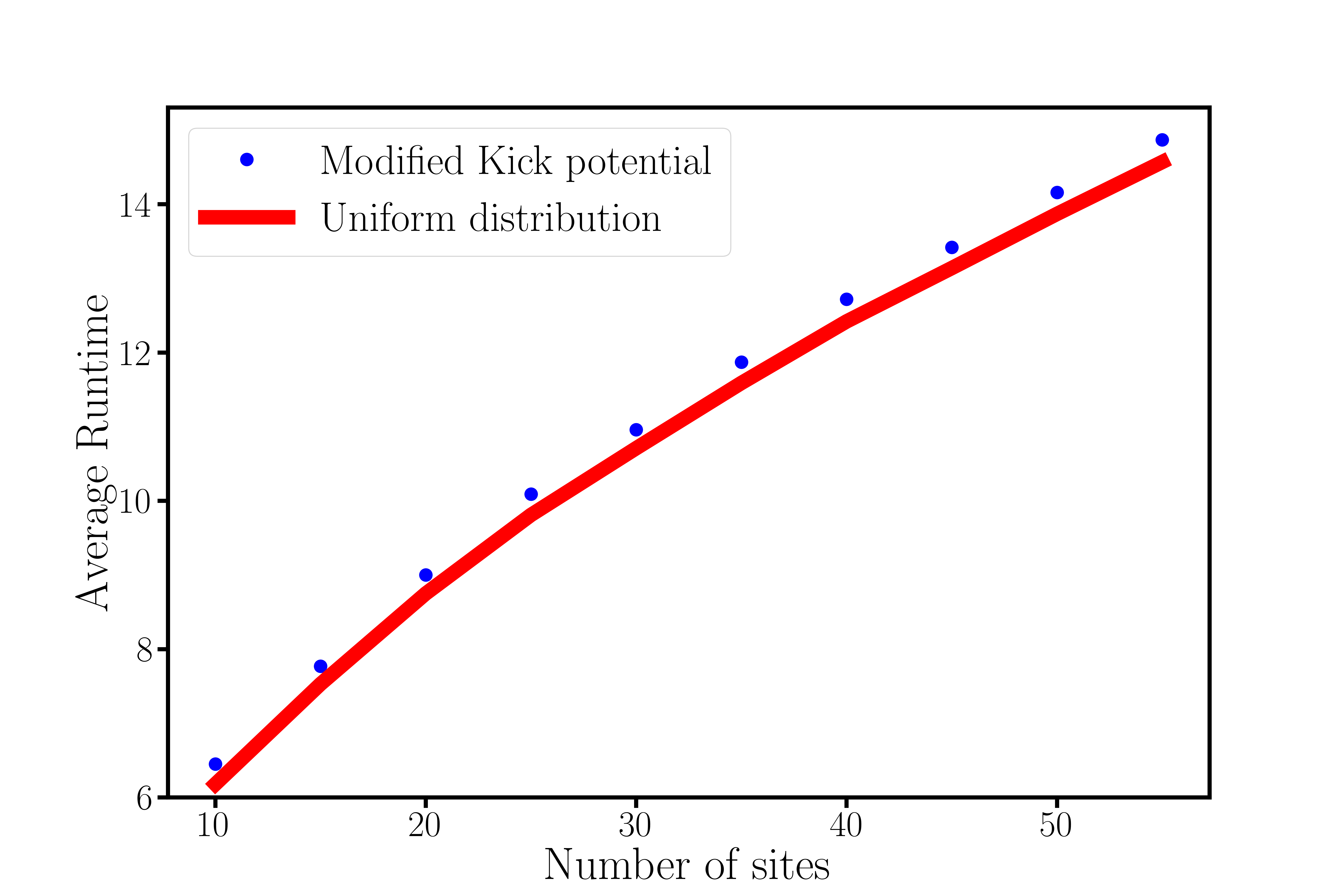}
        \caption{ The average runtime is compared to the one obtained from a uniform distribution.}
    \end{subfigure}
    \caption{Amplitude amplification using the modified kick potential for different numbers of oracle calls. Numerical results indicate that when multiple sites are marked, quadratic speedup can be achieved with $N=\sqrt{3}\sigma$.}
  \label{fig:AA using modified potential} 
\end{figure}

\begin{figure}[t]
    \centering
    \begin{subfigure}{\columnwidth}
        \centering
        \includegraphics[width=\columnwidth]{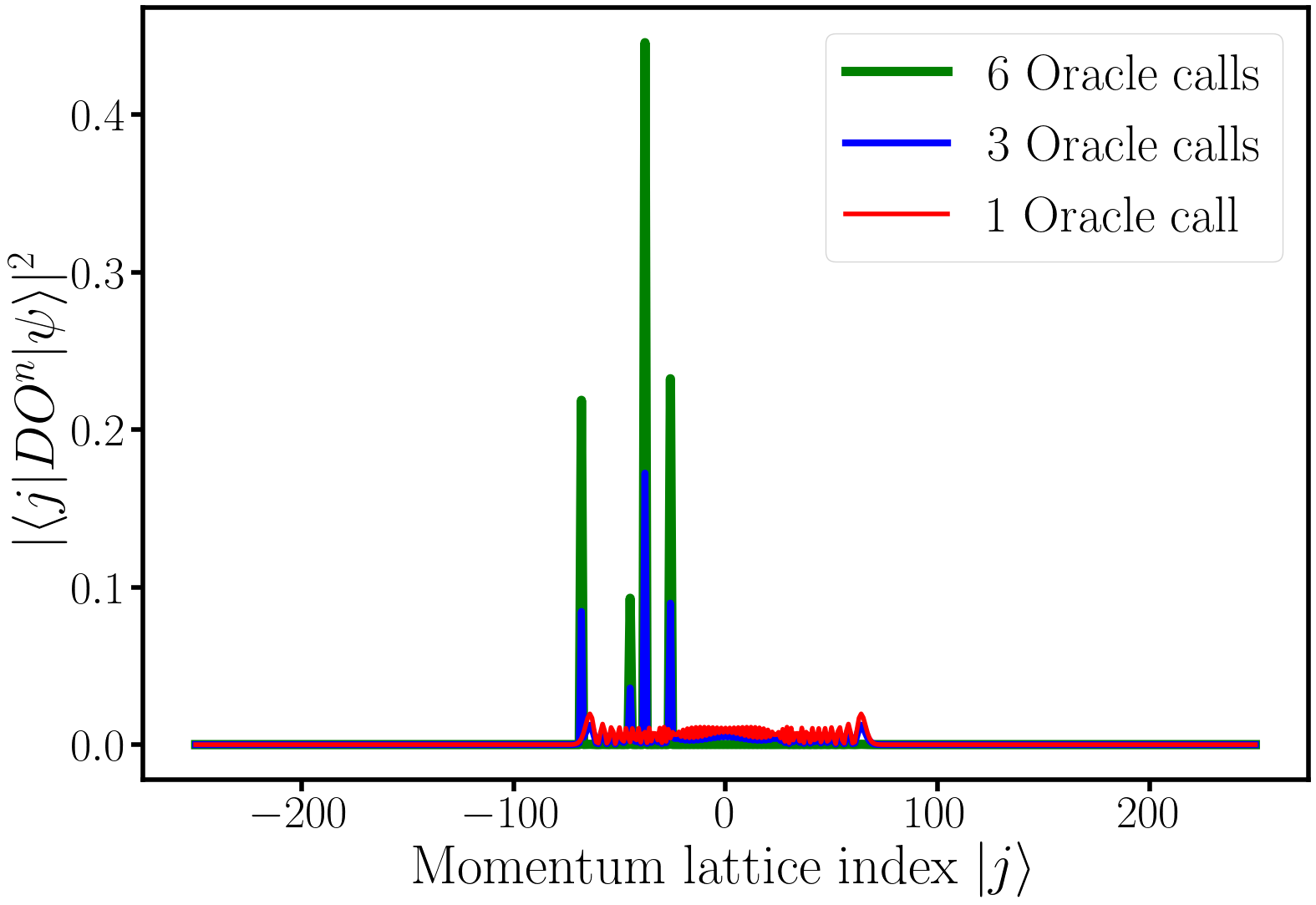}
        \caption{Probability distribution after a different numbers of oracle calls}
    \end{subfigure}
    \hfill
    \begin{subfigure}[b]{1\columnwidth}
        \centering
        \includegraphics[width=1.1\columnwidth]{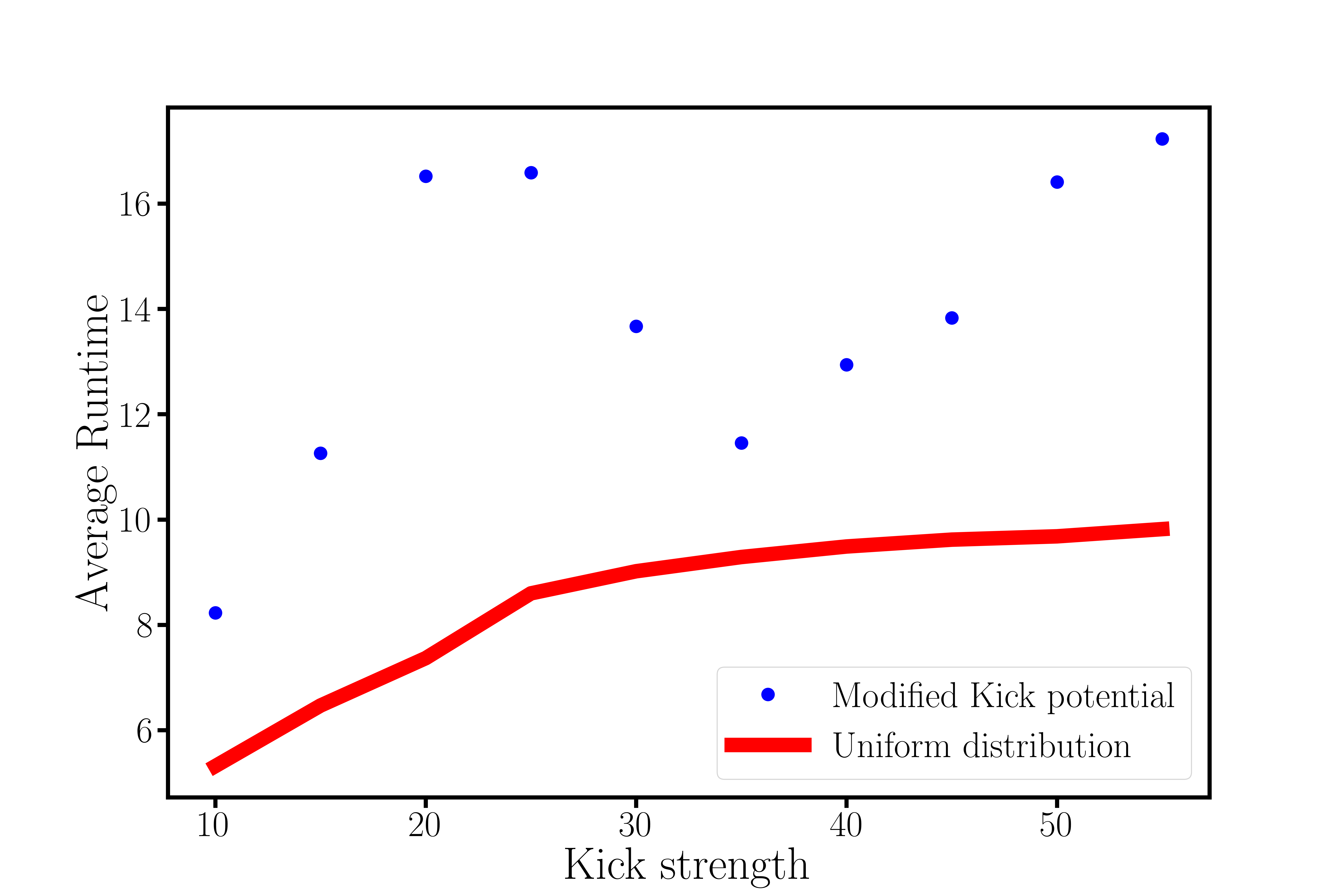}
        \caption{ The average runtime is compared to the one obtained from a uniform distribution.}
    \end{subfigure}
   \caption{Amplitude Amplification procedure visualized using controlled detuning. The initial distribution is improved by applying kicks detuned from resonance.}
  \label{fig:AA using detuned rotor} 
\end{figure}

The expected speedup can be estimated as follows: the initial distribution over momentum sites in Fig. \ref{fig:Density profiles from detuning and M.P compared} is nearly uniform, and let $\sigma$ denote its standard deviation about the mean. Assuming that it truncates sharply beyond $\sqrt{3}\sigma$, (a property of a uniform distribution with standard deviation $\sigma$) the number of search sites is taken to be  $N=2\sqrt{3}\sigma$.
Then, the average runtime can be defined as
\begin{equation}
T_{\text{avg}} = \frac{1}{2\sqrt{3}\sigma} \sum_{i=-\floor{\sqrt{3}\sigma}}^{\ceil{\sqrt{3}\sigma}} \floor*{\frac{\pi}{4\sin^{-1}{\sqrt{a_i}}}}.
\end{equation}
In Figs. \ref{fig:AA using modified potential}(a) and \ref{fig:AA using detuned rotor}(a), we depict the probability distributions in momentum space after 1, 3 and 6 oracle calls. In both cases, the amplitudes of the marked states increase at the expense of those of the unmarked states as expected as the oracle is called. From our simulations with modified potential, shown in Figs. \ref{fig:AA using modified potential}(b), the average runtime for successful search scales as $\sqrt{N}$, and the scaling coincides with the runtime of the Grover algorithm for a uniformly distributed initial state with the same standard deviation. However, in the case of the controlled detuning approach shown in Fig. \ref{fig:AA using detuned rotor}(b), a strict quadratic scaling of the rutime is absent, but the runtimes are much closer to being quadratic than the corresponding runtimes from the standard kick potential with no detuning since there are some sites with very low amplitudes in the latter case. 
\section{Estimating the number of iterations}
\label{sec: Estimating the required number of iterations}
In our implementation of amplitude amplification, the marked sites are unknown, and hence it is necessary to apriori determine the factor $\sqrt{a}(=\sin \frac{\theta}{2})$ in Eq. \ref{iters}. As the initial state in Eq. \ref{eq:instate} is not uniform over the momentum basis and marked site indices are unknown, the required number of iterations $r$ cannot be estimated from the standard deviation. In what follows, we estimate $a_i$ using a variant of Kitaev's phase estimation procedure \cite{kitaev1995quantum,Nielsenchuang}.

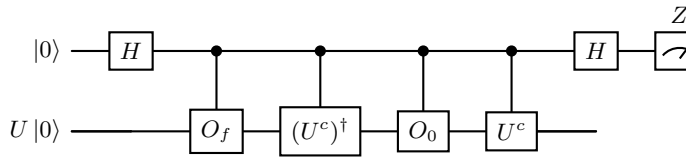
\begin{figure}[b]
\begin{quantikz}
\label{quantikz: Qcirc}
  \lstick{$\ket{0}$} & \gate{H} & \ctrl{1} & \ctrl{1} & \ctrl{1} & \ctrl{1} & \gate{H} &\meter{Z} \\
  \lstick{$U\ket{0}$}    & \qw  & \gate{O_f} & \gate{(U^c)^\dag} & \gate{O_0} &  \gate{U^c}& \qw \\
\end{quantikz}
\caption{Circuit for the phase estimation procedure.}
\end{figure}

The circuit in Fig. \ref{quantikz: Qcirc} represents the entire phase estimation process and its implementation will require a kicked rotor along with an additional spin-1/2 degree of freedom. Firstly, Hadamard gate is applied to the spin-1/2 degree of freedom. Next, the oracle controlled by the spin-$\frac{1}{2}$ state which can be decomposed as follows is implemented
\begin{align*}
O_f^c & = \ket{0}\bra{0}\otimes \mathbb{I} + \ket{1}\bra{1}\otimes O_f
        = \mathbb{I} \otimes \mathbb{I} - 2\ket{1}\bra{1}\otimes P_G,
 \\
      & = \mathbb{I} \otimes P_B + Z\otimes P_G,  \\
      & = \sum_{n\notin M}\mathbb{I} \otimes \ket{n}\bra{n} + \sum_{m\notin M}Z \otimes \ket{m}\bra{m},
\end{align*}
where $M$ is the set of marked momentum states. The controlled version of the $U$ can be expressed as
\begin{multline*}
    U^c = e^{-i (\phi \ket{1}\bra{1})\otimes V(\theta))} = \ket{0}\bra{0}\otimes \mathbb{I} + \ket{1}\bra{1} \otimes e^{-i V(\theta)\phi}.
\end{multline*}
The controlled oracle and the diffusion operator together constitute a single iteration of amplitude amplification. On the subspace spanned by $P_G\ket{\psi}$ and $P_B \ket{\psi}$, the action of a single iteration of amplitude amplification is effectively a rotation on the same plane, whereas for any vector in the subspace orthogonal to it, the amplitude amplification unitary acts as the identity. The state $\ket{\psi}$ can be expressed as a linear combination of two eigenvectors of the amplitude amplification unitary, $\ket{\pm}$, with eigenvalues $-e^{\pm i\theta}$. This yields
\begin{align*}
    \ket{\psi} & = \sin(\frac{\theta}{2})\ket{s} + \cos({\frac{\theta}{2}})\ket{w} =\frac{ie^{\frac{-i\theta}{2}}}{\sqrt{2}}\ket{+} - \frac{ie^{\frac{i\theta}{2}}}{\sqrt{2}}\ket{-},
    \\
    & \ket{s} = \frac{P_G\ket{\psi}}{\bra{\psi}P_G\ket{\psi}}, \;\;\; \ket{w} = \frac{P_B\ket{\psi}}{\bra{\psi}P_B\ket{\psi}}, 
    \\ 
    & \ket{\pm} = \frac{\pm i \ket{s} + \ket{w}}{\sqrt{2}}.
\end{align*}
Finally, the controlled version of amplitude amplification denoted by the operator $A^c$ leads to
\begin{align*}
     \ket{\psi_2} & = A^c(\ket{+}\otimes\ket{\psi}) \\
                  & = A^c \Big( \frac{\ket{0}+\ket{1}}{\sqrt{2}}\Big) \Big(\frac{ie^{\frac{-i\theta}{2}}}{\sqrt{2}}\ket{+} - \frac{ie^{\frac{i\theta}{2}}}{\sqrt{2}}\ket{-}\Big)
    \\
    & = \frac{ie^{\frac{-i\theta}{2}}}{2} \left( \ket{0}\ket{+} + \ket{1}\ket{-} \right) -
        \frac{i e^{\frac{i\theta}{2}}}{2}  \left( \ket{0}\ket{-} + \ket{1}\ket{+} \right).   
\end{align*}
The expected value of the measurement in the $Z$ basis is $\bra{\psi_2}X\ket{\psi_2} = -\cos(\theta)$ where $\theta$ is the angle required to estimate $r$. Thus, by iterating $r$ times, it is possible to estimate $\theta$ to any desired level of precision. Ideally, this should be done prior to the search algorithm to obtain the necessary number of iterations.
\\
In experiments, the upper wire in Fig. \ref{quantikz: Qcirc} can represent an internal hyperfine state while the lower one represents the states on the angular momentum lattice. The Hadamard gate on these states can be realized using microwave pulses as described in \cite{Dadras_2018}. The oracle controlled by the internal hyperfine state can be implemented using phase gates on the internal state controlled by the momentum-state instead. If the momentum state is one of the marked ones, the phase gate will be implemented on the internal hyperfine level. The momentum-state controlled representation of the controlled oracle indicates that this operation can be experimentally implemented using velocity-selective transitions \cite{PhysRevA.65.013403}, which may be easier to achieve than operations controlled by the internal hyperfine state. The controlled unitaries $U^c$ and $(U^c)^\dag$ can be implemented as special cases of biased DTQWs \cite{PhysRevA.99.043617} with bias 1 for the internal state $\ket{1}$. Since biased DTQWs have already been realized \cite{PhysRevA.99.043617}, this step should be experimentally feasible as well.

\section{Sensitivity to Noisy kicks}
\label{sec: Sensitivity to Noisy kicks}
Imperfections in the kicked rotor pulse sequence which may occur in an experimental realization of the system can be modelled by adding noise to the kicks. It is known that adding a stochastic term to the Hamiltonian of a quantum system can lead to a decay in purity. The classical noise processes can be used to simulate decoherence by means of pure dephasing  \cite{10.1063/1.5099499}. This can be implemented in QKR with kick strength treated as a {\it i.i.d} random variable sampled from a normal distribution $\phi_j\sim N(\phi,\delta)$, where $\phi$ is the mean kick strength, and $\delta$ is the standard deviation. The sensitivity to noise in the kick strength can be understood by studying how the errors scale with the noise-strength $\delta$.

\subsection{Noise-averaged dynamics}
\begin{figure}[t]
  \centering
  \includegraphics[width=\columnwidth]{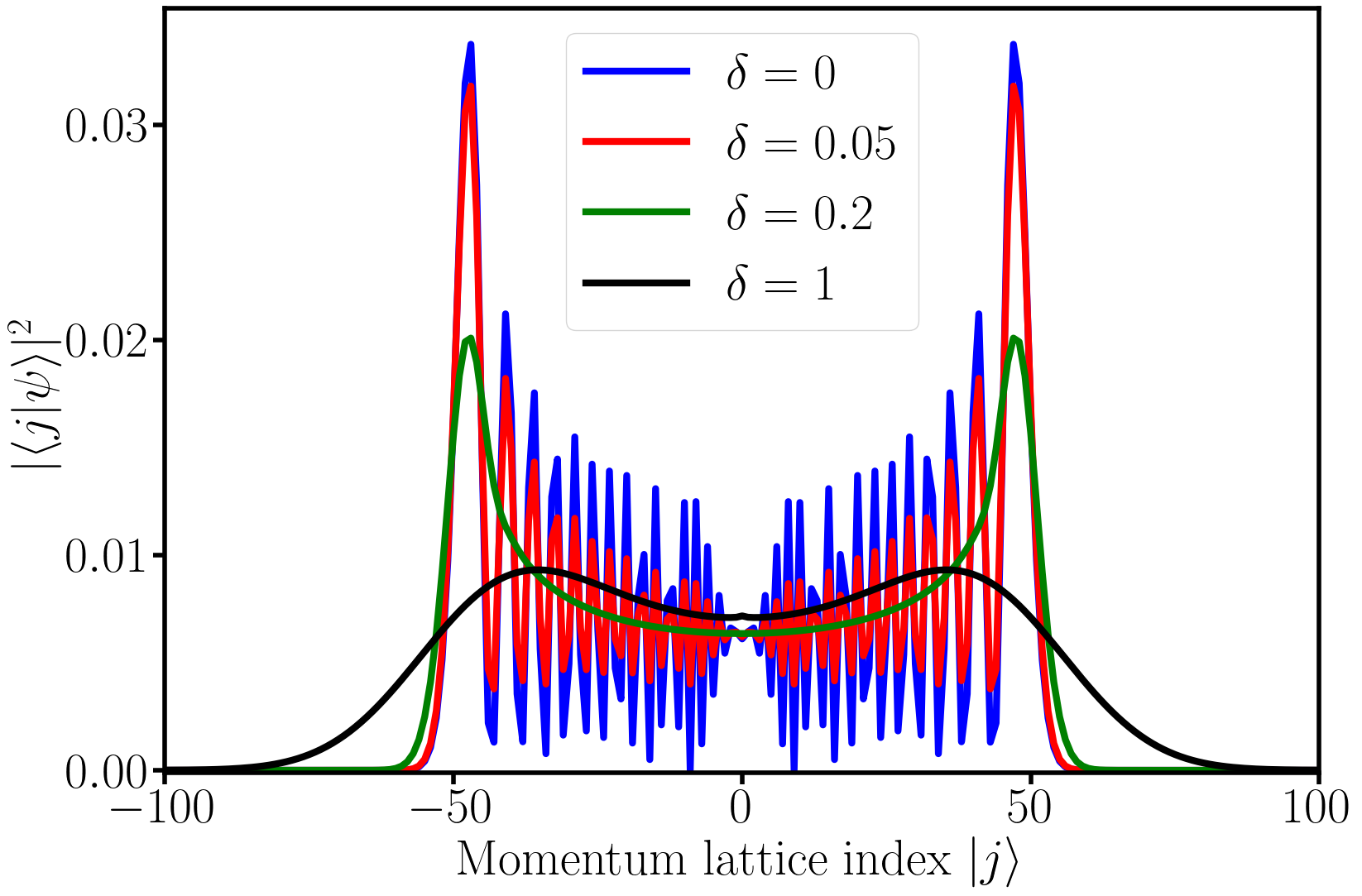}
  \caption{Noise-averaged probability density profiles for the resonant noisy QKR. The mean kick strength was chosen to be $\phi=0.25$ and 200 kicks are performed.}
  \label{fig:NAPD QKR}
\end{figure}

The noise-averaged density operator can reproduce the dynamics of expectation values of observables after multiple realizations of the noise, since trace and noise-averaging are commuting linear operations: $\overline{\Tr{O\rho}} = \Tr{O\bar{\rho}}$, where the overbar represents noise averaging. For a sequence of $m$ noisy kicks, state of the system is
\begin{equation}
    \rho(mT^+)= e^{-i \Tilde{\phi} V(\theta)} \rho(0) e^{i\Tilde{\phi} V(\theta)}; \:\;\;\; \Tilde{\phi} = \sum_{j=1}^m \phi_j. 
\end{equation}    
In position representation, we get    
    %\sim N(mK_0, \sqrt{m}\delta)
\begin{multline}
    \\ \bra{\theta} \rho(mT^+) \ket{\theta'} =  e^{-i \Tilde{\phi} (V(\theta)-V(\theta^{'})} \bra{\theta} \rho(0) \ket{\theta'},
    \\ \bra{\theta} \overline{\rho(mT^+)} \ket{\theta'} = e^{-im\phi(V(\theta)-V(\theta^{'})} F^m(\theta, \theta^{'}) \bra{\theta} \rho(0) \ket{\theta'}, \\
    F(\theta, \theta^{'})=e^{-\frac{\delta^2(V(\theta')-V(\theta))^2}{2}}.
    \label{eq:narho}
\end{multline} 
This is similar to a scenario of adding continuous-time stochastic process to an Hamiltonian, which leads to non-unitary evolution of the ensemble-averaged density operator \cite{article}. 
Now, Eq. \ref{eq:narho} can be used to compute trace with any observable, yielding the corresponding noise-averaged expectation value. From Eq. \ref{eq:narho}, it is also clear that $ \bra{\theta} \rho(mT^+) \ket{\theta'}=0$ for all $\theta\neq \theta'$ and $\theta\neq 2\pi-\theta'$, provided $V(\theta)$ is invertible in $[0,2\pi]$. Thus, most of the off-diagonal matrix elements decay exponentially with time indicating decoherence. The noise-averaged dynamics of $\rho$ does not differ from the noiseless case to first order (as seen by Taylor expanding the noise term in Eq. \ref{eq:narho}). But the second-order correction and all the even-order corrections are non-zero and can have a significant impact on the probability distribution for large noise amplitudes. The momentum space probability density profile for QKR with $V(\theta) = \cos \theta$ subjected to a few noisy kicks is depicted in Fig. \ref{fig:NAPD QKR}.

\subsection{Effects of noisy kicks on Amplitude Amplification}
\begin{figure}[t]
  \centering
  \includegraphics[width=\columnwidth]{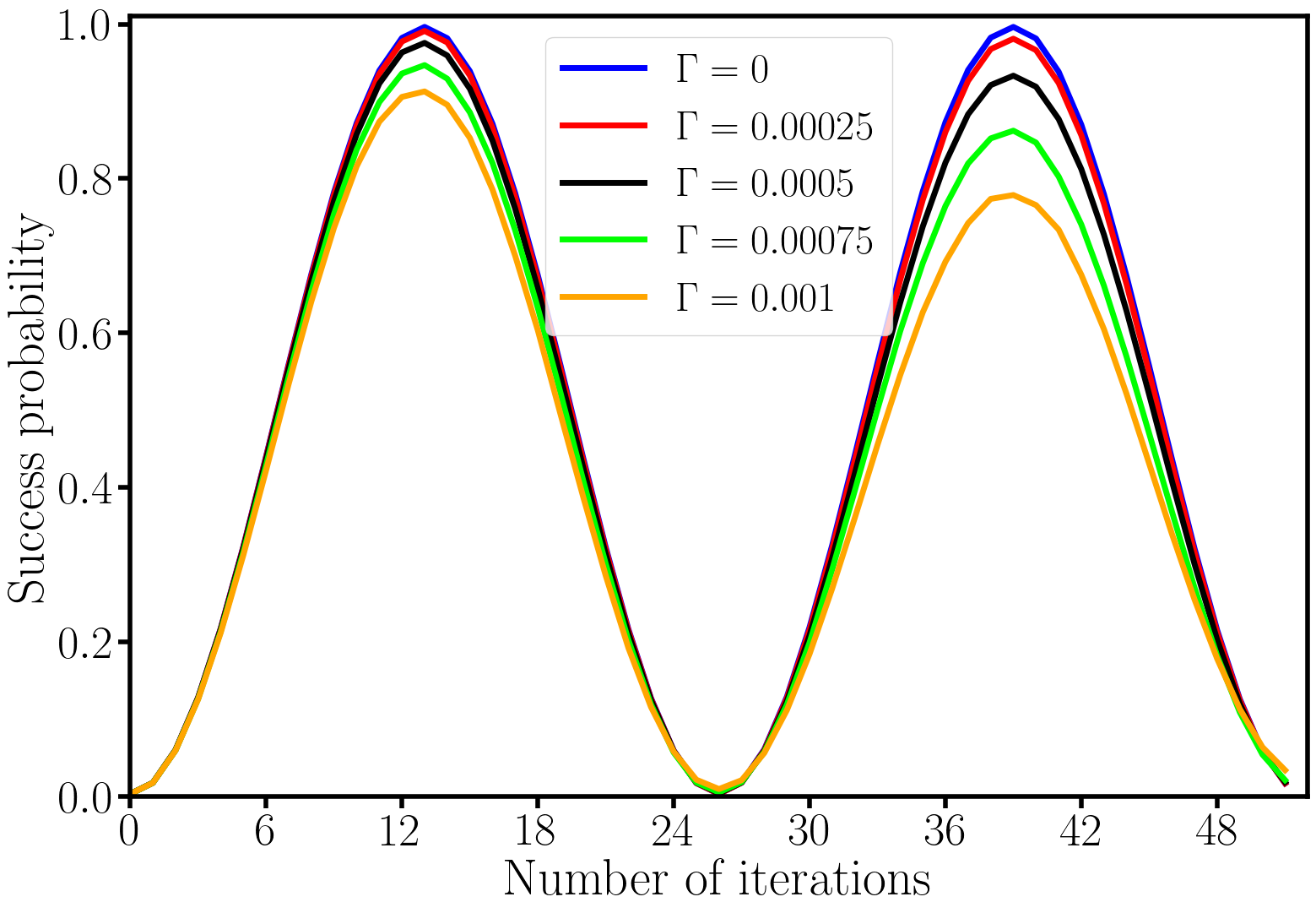}
  \caption{The plots depict noise-averaged success probabilities after various numbers of iterations of amplitude amplification. The noise strength reported is normalized by the kick strength}
  \label{fig:NA succes probabilities}
\end{figure}

 In the noiseless case, the success probability is an oscillating sinusoidal function of the number of iterations of the amplitude amplification. As Fig. \ref{fig:NA succes probabilities} shows, the noise-averaged success probability decreases as the noise strength $\Gamma = \delta/\phi$ increases, and the fidelity drops. Averaging over multiple noise realizations is necessary to smoothen the irregular fluctuations.

If a closed-form for evolution of the noise-averaged density operator is known, then the noise-averaged fidelity could be obtained by taking the trace with $P_G = \sum_{m\in M}\ket{m}\bra{m}$. The state at $r+1^{\rm th}$ iteration is related to the state after $r^{\rm th}$ iteration through
\begin{multline*}
    \bra{\theta}\rho^{r+1}\ket{\theta'}  =  \bra{\theta}D^rO\rho^{r}O(D^r)^\dag\ket{\theta'},
    \\
    D^r= e^{-i\delta\phi_{2r}V(\theta)}(\mathbb{I}-2\ket{\psi_0}\bra{\psi_0})e^{i\delta\phi_{2r-1}V(\theta)},
\end{multline*}
where $\delta\phi_{l} = \phi_l-\phi$ are independent random variables for all $0\leq l \leq 2r$. Hence, to obtain full noise averaged result, independently averaging over $\delta\phi_{l}$ is possible. This can also be shown by explicitly calculating the first order correction due to the noise. Consider the state vector after $N$ iterations of amplitude amplification :
\begin{multline*}
    \ket{\psi^N} = \Pi_{r=1}^N[e^{-i\delta N_{2r} V(\theta)}(\mathbb{I}-2\ket{\psi_0}\bra{\psi_0}) 
    \\
    e^{i\delta N_{2r-1}V(\theta)} (\mathbb{I}-2P_G)]\ket{\psi}.
\end{multline*}
We use the fact that $\delta\phi_l =\delta N_l \sim \delta N(0,1)$ since $N_l$ is a normal random variable. Now, by expanding the state vector in the angular momentum eigenbasis $\ket{\psi^N} = \sum_{n\in\mathbb{Z}}c_n(\delta)\ket{n}$, the derivative of $c_n$ with respect to $\delta$, at $\delta=0$ can be expressed as
\begin{multline*}
\frac{dc_n(\delta)}{d\delta} = i \sum_{r=1}^N\bra{n} G^{N-r}\Big[-\phi_{2r}V(\theta)G + 
\\
\phi_{2r-1}(\mathbb{I}-2\ket{\psi_0}\bra{\psi_0})V(\theta)(\mathbb{I}- 2P_G)  \Big]G^{r-1}\ket{\psi}.
\end{multline*}
However, $\langle N_l \rangle = 0$ for all $0 \leq l \leq 2N$. Thus, the noise-averaged first order correction (in the noise strength $\delta$) to the success probability vanishes. On the other hand, the second order correction is non-zero. On similar grounds, it can be argued that all odd-order corrections would evaluate to zero, since all odd moments of the variables $N_l$ are zero, leaving only even-order corrections.

\begin{figure}[b]
  \centering
  \includegraphics[width=\columnwidth]{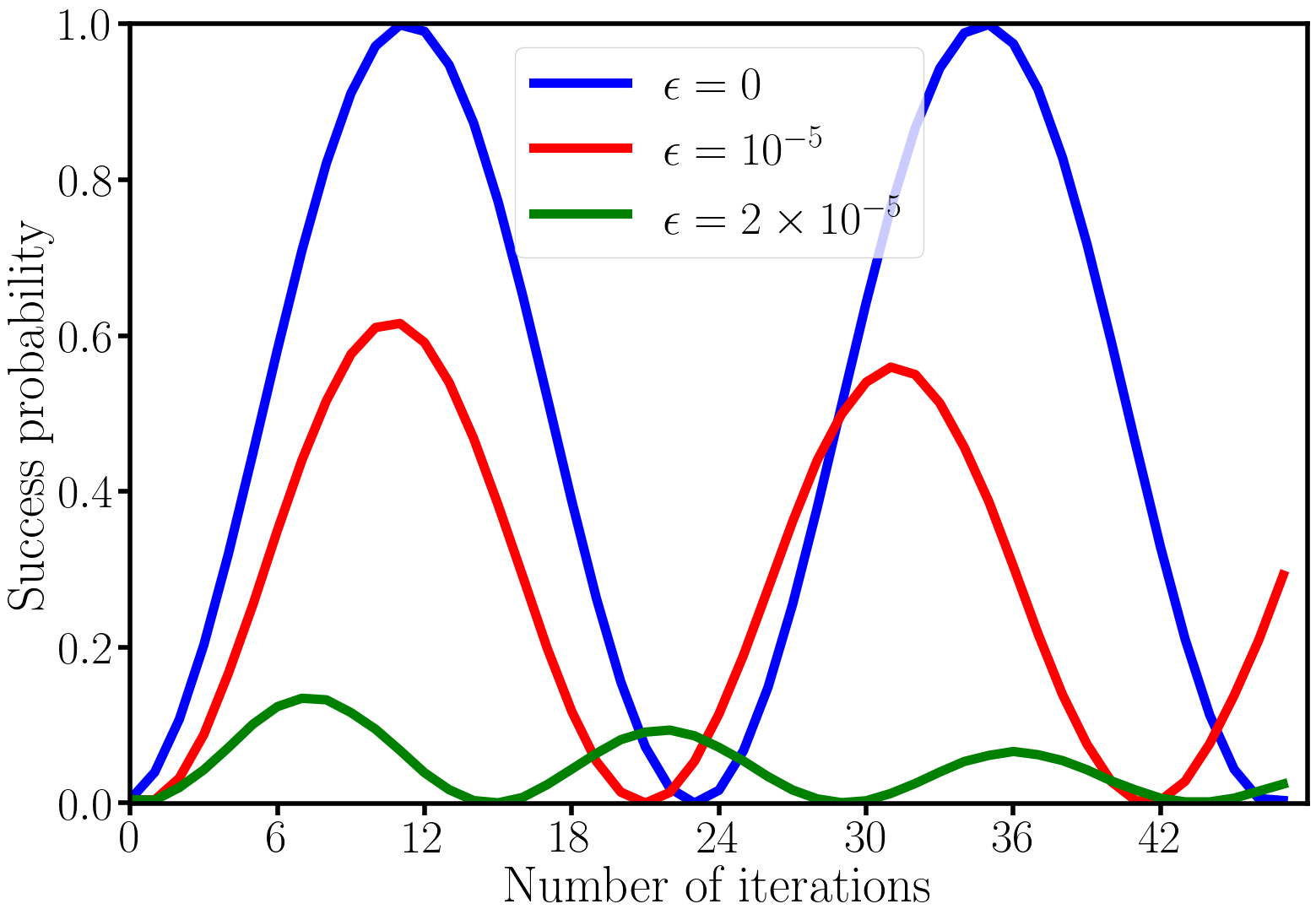}
  \caption{Success probability as a function of detuning from resonance(scaled by the kick period) denoted by $\epsilon = \epsilon'/T$. Fidelity very quickly decays as with increasing $\epsilon$.}
  \label{fig:Success probabilities for erroneous detuning from resonance}
\end{figure}

Since the noise-averaged corrections scale as $~\delta^2$ to the lowest order, our Amplitude Amplification procedure is reasonably robust to noise in the kick strengths.  Fig. \ref{fig:NA succes probabilities} indicates that the maximum fidelity decays with an increase in rescaled noise strength $\Gamma = \delta/\phi$. However, sensitivity to noise is much lower than the sensitivity to errors in detuning, where the latter are rescaled by the kick period. We see from Fig. \ref{fig:NA succes probabilities} and Fig. \ref{fig:Success probabilities for erroneous detuning from resonance} that comparable values of fidelity occur at $\Gamma\sim 10^{-3}$ and rescaled detuning $\epsilon\sim 10^{-5}$ indicating that the Amplitude Amplification procedure is much more robust to noisy kicks than a fixed detuning from resonance.

\begin{figure}[t]
  \includegraphics[width=\columnwidth]{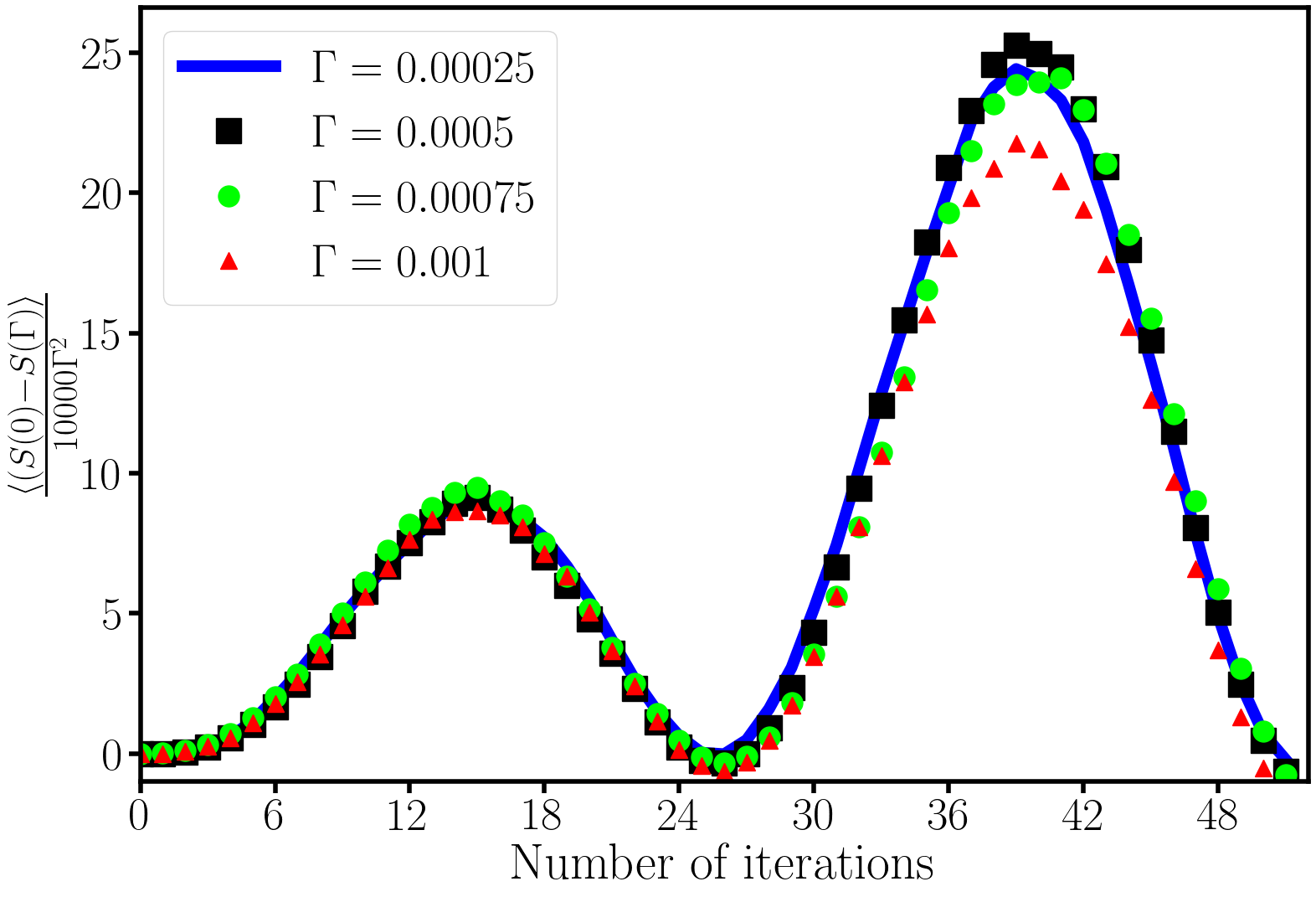}
  \caption{Deviations of the success probability curves $S(\Gamma)$ (averaging over multiple noise realizations) for various noise strengths $\Gamma$ from the noiseless curve $S(\Gamma=0)$, rescaled by a factor $10^4\Gamma^2$ have been plotted as functions of number of iterations.}
  \label{fig:Scaling of errors with noise strength}
\end{figure}

The deviations in the success probability of the noiseless case, rescaled by $\Gamma^2$, are displayed in Fig.  \ref{fig:Scaling of errors with noise strength}. These approximate collapse of all the curves indicates that the quadratic scaling prediction holds true. We note that the deviations from the noiseless cases are vanishingly small though it appears large due to scaling by $\Gamma^2$. The extra factor of $10^4$ in Fig. \ref{fig:Scaling of errors with noise strength} is chosen so the that deviations are $\order{1}$, while the rescaled noise strengths $\Gamma$ are of $\order{10^{-4}}-\order{10^{-3}}$.

%\subsection{}
%\begin{figure}[t]
%\begin{center}
%\includegraphics*[scale=0.325]{}
%\caption{(Color online)}
%\label{fig:phasespaceplot1}
%end{center}
%\end{figure}
%\begin{figure}[h!]
%\begin{center}
%\includegraphics*[scale=0.80]{}
%\includegraphics*[scale=0.80]{}
%\caption{}
%\label{fig:jis1Entropy}
%\end{center}
%\end{figure}
%\subsection{}
% Thus, it can be seen easily that the dimension of $\widehat{O}_0$ 
% and number of times $c_0$ comes both are equal to one. Similarly, the dimension of $\widehat{O}_1$ and number of 
% times $c_1$ comes both are equal to $2j \choose 1$.

% \begin{figure}[t]
% \begin{center}
% \includegraphics[scale=0.5]{temp5.eps}
% \includegraphics[scale=0.5]{temp5.eps}
% \caption{(Color online) $j=3$}
% \end{center}
% \end{figure}
% 
% \begin{figure}[t!]
% \begin{center}
% \includegraphics*[scale=0.5]{temp5.eps}
% \includegraphics*[scale=0.5]{temp5.eps}
% \caption{(Color online) $j=3$}
% \end{center}
% \end{figure}
% 
% \begin{figure}[t!]
% \begin{center}
% \includegraphics*[scale=0.5]{temp5.eps}
% \includegraphics*[scale=0.5]{temp5.eps}
% \caption{(Color online) $j=3$}
% \end{center}
% \end{figure}
% \begin{figure}[t!]
% \begin{center}
% \includegraphics*[scale=0.5]{temp5.eps}
% \includegraphics*[scale=0.5]{temp5.eps}
% \caption{(Color online) $j=3$}
% \end{center}
% \end{figure}
\section{Conclusions and outlook}
\label{sec: Conclusions and outlook}
We have presented a scheme for amplitude amplification using standard atom-optics manipulation techniques which can be achieved using a quantum kicked rotor model with an internal spin-$\frac{1}{2}$ degree of freedom. The amplitude amplification algorithm requires three ingredients: (a) a method to implement a unitary operator $U$ not diagonal in the basis of interest, (b) a method for implementing time-reversal of the dynamics caused by $U$, (c) a method to obtain the phase kickback required for the Grover oracle. All of these unitary operations can be performed on the kicked rotor, which makes it a suitable platform for the implementation of the algorithm through atom-optics experiments. Thus, the atom-optics kicked rotor is a platform useful for more than just phenomenological studies of quantum chaos and Anderson localization.
\\
%Furthermore, we have shown that the system's characteristic property of dynamical localization in momentum space can be used to our advantage, to engineer a momentum-space probability distribution more conducive to amplitude amplification than the one produced by a momentum-space continuous-time quantum walk. 
In order to obtain an initial state uniformly distributed over momentum basis, we employ a modified kick potential which allows for long-range hopping. Further, a new scheme for amplitude estimation for the quantum kicked rotor has been proposed. Finally, we study the effects of noise on the kick strengths. Since the variant of kicked rotor considered here is not quantum chaotic, it is of interest to study if a quantum chaotic system can be used for algorithmic purposes, and whether this will have any advantage over other platforms for computation.

Our proposals are rooted in viable techniques already realized in experiments. The periodic boundary conditions for the spatial degree of freedom could be achieved using a toroidal non-interacting BEC \cite{InteratingQKRs}. The $\pi$ phase shifts required for marking are performed by exploiting velocity-selective Raman transitions to select certain momentum values corresponding to the lattice sites to be marked and taking the internal 2-level system on an adiabatic cycle which leads to a global geometric phase of $\pi$ for the internal state. However, this operation is done selectively on certain momentum states, and becomes a relative phase for these alone. 

Realizing an arbitrary kick potential is certainly possible from the point of view of an experiment since each term in the Fourier series corresponds to a different laser with a different wavelength. An operation controlled by the internal state in phase estimation is required, and such operations might typically be challenging to perform. However, we have showed that this operation can be rewritten in terms of momentum-selective phase gates which may once again be implemented using velocity-selective Raman transitions. We expect that the results in this work would lead to experimental realization of amplitude amplification using atom-optics test beds.
\acknowledgments
We thank S. Sagar Maurya for many discussions about the experimental feasibility of the procedures proposed as well as suggestions on other aspects of this work. We also thank J Bharati Kannan for some suggestions to improve the fidelity of the search protocol. We also thank Professor T.S. Mahesh for useful discussion about an oracle which makes use of a geometric phase, as well as phase estimation using Kitaev's procedure. 

\appendix 
\section{Detuning from resonance}
\label{sec: Effects of erroneous detuning from resonance}
The QKR system is tuned to resonance for two reasons: Firstly, a resonant kicked rotor will not exhibit dynamical localization. Instead, the momentum-space wave-packet displays a ballistic spread, and a superposition is created fairly quickly. In the present case, the mean energy grows quadratically with time; $\langle E \rangle \sim n^2$. The second advantage is that resonant dynamics can be reversed by simply shifting the phase of the kick potential. However,  even a small (uncontrolled) detuning from resonance will make the dynamics irreversible. This might be relevant if the marking process employed is slow, in which case the undesired free evolution of QKR during the marking step must also be accounted for. The forward and backward kick sequences are given by
\[
U_f= e^{\frac{-iL^2\epsilon'}{2I\hbar}}e^{-i\phi \cos(\theta)}, \;\;\:\;\; U_b = e^{\frac{-iL^2\epsilon'}{2I\hbar}}e^{i\phi \cos(\theta)},
\]
where $\epsilon'$ represents detuning from quantum resonance.
Each time a kick is imparted, the state is rotated away from the desired state by a small angle due to the free evolution part of the Floquet operator. In this scenario, even if the kick strength alternates between positive and negative values, it is impossible to realize the diffusion operator which requires time-reversed evolution in the form $U_f^\dag \neq U_b$. Detuning from resonance leads to errors that build up after each Grover iteration, and could lead to large cumulative errors. This is depicted in Fig. \ref{fig:Success probabilities for erroneous detuning from resonance}, where we report success probability for detuning $\epsilon'$ rescaled by the kick period $T$, so that the relevant time-scale is $\epsilon=\epsilon'/T$. This analysis suggests that the effects of detuning from quantum resonance in an QKR-Amplitude Amplification experiment may be quite drastic.

\bibliographystyle{apsrev4-2}
\bibliography{sample}

\begin{thebibliography}{40}%
\makeatletter
\providecommand \@ifxundefined [1]{%
 \@ifx{#1\undefined}
}%
\providecommand \@ifnum [1]{%
 \ifnum #1\expandafter \@firstoftwo
 \else \expandafter \@secondoftwo
 \fi
}%
\providecommand \@ifx [1]{%
 \ifx #1\expandafter \@firstoftwo
 \else \expandafter \@secondoftwo
 \fi
}%
\providecommand \natexlab [1]{#1}%
\providecommand \enquote  [1]{``#1''}%
\providecommand \bibnamefont  [1]{#1}%
\providecommand \bibfnamefont [1]{#1}%
\providecommand \citenamefont [1]{#1}%
\providecommand \href@noop [0]{\@secondoftwo}%
\providecommand \href [0]{\begingroup \@sanitize@url \@href}%
\providecommand \@href[1]{\@@startlink{#1}\@@href}%
\providecommand \@@href[1]{\endgroup#1\@@endlink}%
\providecommand \@sanitize@url [0]{\catcode `\\12\catcode `\$12\catcode `\&12\catcode `\#12\catcode `\^12\catcode `\_12\catcode `\%12\relax}%
\providecommand \@@startlink[1]{}%
\providecommand \@@endlink[0]{}%
\providecommand \url  [0]{\begingroup\@sanitize@url \@url }%
\providecommand \@url [1]{\endgroup\@href {#1}{\urlprefix }}%
\providecommand \urlprefix  [0]{URL }%
\providecommand \Eprint [0]{\href }%
\providecommand \doibase [0]{https://doi.org/}%
\providecommand \selectlanguage [0]{\@gobble}%
\providecommand \bibinfo  [0]{\@secondoftwo}%
\providecommand \bibfield  [0]{\@secondoftwo}%
\providecommand \translation [1]{[#1]}%
\providecommand \BibitemOpen [0]{}%
\providecommand \bibitemStop [0]{}%
\providecommand \bibitemNoStop [0]{.\EOS\space}%
\providecommand \EOS [0]{\spacefactor3000\relax}%
\providecommand \BibitemShut  [1]{\csname bibitem#1\endcsname}%
\let\auto@bib@innerbib\@empty
%</preamble>
\bibitem [{\citenamefont {Brassard}\ \emph {et~al.}(2002)\citenamefont {Brassard}, \citenamefont {H\o~yer}, \citenamefont {Mosca},\ and\ \citenamefont {Tapp}}]{Brassard_2002}%
  \BibitemOpen
  \bibfield  {author} {\bibinfo {author} {\bibfnamefont {G.}~\bibnamefont {Brassard}}, \bibinfo {author} {\bibfnamefont {P.}~\bibnamefont {H\o~yer}}, \bibinfo {author} {\bibfnamefont {M.}~\bibnamefont {Mosca}},\ and\ \bibinfo {author} {\bibfnamefont {A.}~\bibnamefont {Tapp}},\ }in\ \href {https://doi.org/10.1090/conm/305/05215} {\emph {\bibinfo {booktitle} {Quantum computation and information ({W}ashington, {DC}, 2000)}}},\ \bibinfo {series} {Contemp. Math.}, Vol.\ \bibinfo {volume} {305}\ (\bibinfo  {publisher} {Amer. Math. Soc., Providence, RI},\ \bibinfo {year} {2002})\ pp.\ \bibinfo {pages} {53--74}\BibitemShut {NoStop}%
\bibitem [{\citenamefont {H\o{}yer}(2000)}]{Hoyer_2000}%
  \BibitemOpen
  \bibfield  {author} {\bibinfo {author} {\bibfnamefont {P.}~\bibnamefont {H\o{}yer}},\ }\href {https://doi.org/10.1103/PhysRevA.62.052304} {\bibfield  {journal} {\bibinfo  {journal} {Phys. Rev. A}\ }\textbf {\bibinfo {volume} {62}},\ \bibinfo {pages} {052304} (\bibinfo {year} {2000})}\BibitemShut {NoStop}%
\bibitem [{\citenamefont {Grover}(1997)}]{grover1996fast}%
  \BibitemOpen
  \bibfield  {author} {\bibinfo {author} {\bibfnamefont {L.~K.}\ \bibnamefont {Grover}},\ }\href {https://doi.org/10.1103/PhysRevLett.79.325} {\bibfield  {journal} {\bibinfo  {journal} {Phys. Rev. Lett.}\ }\textbf {\bibinfo {volume} {79}},\ \bibinfo {pages} {325} (\bibinfo {year} {1997})}\BibitemShut {NoStop}%
\bibitem [{\citenamefont {Grover}(1998)}]{PhysRevLett.80.4329}%
  \BibitemOpen
  \bibfield  {author} {\bibinfo {author} {\bibfnamefont {L.~K.}\ \bibnamefont {Grover}},\ }\href {https://doi.org/10.1103/PhysRevLett.80.4329} {\bibfield  {journal} {\bibinfo  {journal} {Phys. Rev. Lett.}\ }\textbf {\bibinfo {volume} {80}},\ \bibinfo {pages} {4329} (\bibinfo {year} {1998})}\BibitemShut {NoStop}%
\bibitem [{\citenamefont {Chuang}\ \emph {et~al.}(1998)\citenamefont {Chuang}, \citenamefont {Gershenfeld},\ and\ \citenamefont {Kubinec}}]{ChuGerKub1998}%
  \BibitemOpen
  \bibfield  {author} {\bibinfo {author} {\bibfnamefont {I.~L.}\ \bibnamefont {Chuang}}, \bibinfo {author} {\bibfnamefont {N.}~\bibnamefont {Gershenfeld}},\ and\ \bibinfo {author} {\bibfnamefont {M.}~\bibnamefont {Kubinec}},\ }\href {https://doi.org/10.1103/PhysRevLett.80.3408} {\bibfield  {journal} {\bibinfo  {journal} {Phys. Rev. Lett.}\ }\textbf {\bibinfo {volume} {80}},\ \bibinfo {pages} {3408} (\bibinfo {year} {1998})}\BibitemShut {NoStop}%
\bibitem [{\citenamefont {Das}\ \emph {et~al.}(2003)\citenamefont {Das}, \citenamefont {Mahesh},\ and\ \citenamefont {Kumar}}]{DAS20038}%
  \BibitemOpen
  \bibfield  {author} {\bibinfo {author} {\bibfnamefont {R.}~\bibnamefont {Das}}, \bibinfo {author} {\bibfnamefont {T.}~\bibnamefont {Mahesh}},\ and\ \bibinfo {author} {\bibfnamefont {A.}~\bibnamefont {Kumar}},\ }\href {https://doi.org/https://doi.org/10.1016/S0009-2614(02)01895-X} {\bibfield  {journal} {\bibinfo  {journal} {Chemical Physics Letters}\ }\textbf {\bibinfo {volume} {369}},\ \bibinfo {pages} {8} (\bibinfo {year} {2003})}\BibitemShut {NoStop}%
\bibitem [{\citenamefont {Brickman}\ \emph {et~al.}(2005)\citenamefont {Brickman}, \citenamefont {Haljan}, \citenamefont {Lee}, \citenamefont {Acton}, \citenamefont {Deslauriers},\ and\ \citenamefont {Monroe}}]{BriHalLee2005}%
  \BibitemOpen
  \bibfield  {author} {\bibinfo {author} {\bibfnamefont {K.-A.}\ \bibnamefont {Brickman}}, \bibinfo {author} {\bibfnamefont {P.~C.}\ \bibnamefont {Haljan}}, \bibinfo {author} {\bibfnamefont {P.~J.}\ \bibnamefont {Lee}}, \bibinfo {author} {\bibfnamefont {M.}~\bibnamefont {Acton}}, \bibinfo {author} {\bibfnamefont {L.}~\bibnamefont {Deslauriers}},\ and\ \bibinfo {author} {\bibfnamefont {C.}~\bibnamefont {Monroe}},\ }\href {https://doi.org/10.1103/PhysRevA.72.050306} {\bibfield  {journal} {\bibinfo  {journal} {Phys. Rev. A}\ }\textbf {\bibinfo {volume} {72}},\ \bibinfo {pages} {050306} (\bibinfo {year} {2005})}\BibitemShut {NoStop}%
\bibitem [{\citenamefont {Figgatt}\ \emph {et~al.}(2017)\citenamefont {Figgatt}, \citenamefont {Maslov}, \citenamefont {Landsman}, \citenamefont {Linke}, \citenamefont {Debnath},\ and\ \citenamefont {Monroe}}]{Figgatt_2017}%
  \BibitemOpen
  \bibfield  {author} {\bibinfo {author} {\bibfnamefont {C.}~\bibnamefont {Figgatt}}, \bibinfo {author} {\bibfnamefont {D.}~\bibnamefont {Maslov}}, \bibinfo {author} {\bibfnamefont {K.~A.}\ \bibnamefont {Landsman}}, \bibinfo {author} {\bibfnamefont {N.~M.}\ \bibnamefont {Linke}}, \bibinfo {author} {\bibfnamefont {S.}~\bibnamefont {Debnath}},\ and\ \bibinfo {author} {\bibfnamefont {C.}~\bibnamefont {Monroe}},\ }\bibfield  {journal} {\bibinfo  {journal} {Nature Communications}\ }\textbf {\bibinfo {volume} {8}},\ \href {https://doi.org/10.1038/s41467-017-01904-7} {10.1038/s41467-017-01904-7} (\bibinfo {year} {2017})\BibitemShut {NoStop}%
\bibitem [{\citenamefont {DiCarlo}\ \emph {et~al.}(2009)\citenamefont {DiCarlo}, \citenamefont {Chow}, \citenamefont {Gambetta}, \citenamefont {Bishop}, \citenamefont {Johnson}, \citenamefont {Schuster}, \citenamefont {Majer}, \citenamefont {Blais}, \citenamefont {Frunzio}, \citenamefont {Girvin} \emph {et~al.}}]{DicChoGam2009}%
  \BibitemOpen
  \bibfield  {author} {\bibinfo {author} {\bibfnamefont {L.}~\bibnamefont {DiCarlo}}, \bibinfo {author} {\bibfnamefont {J.~M.}\ \bibnamefont {Chow}}, \bibinfo {author} {\bibfnamefont {J.~M.}\ \bibnamefont {Gambetta}}, \bibinfo {author} {\bibfnamefont {L.~S.}\ \bibnamefont {Bishop}}, \bibinfo {author} {\bibfnamefont {B.~R.}\ \bibnamefont {Johnson}}, \bibinfo {author} {\bibfnamefont {D.}~\bibnamefont {Schuster}}, \bibinfo {author} {\bibfnamefont {J.}~\bibnamefont {Majer}}, \bibinfo {author} {\bibfnamefont {A.}~\bibnamefont {Blais}}, \bibinfo {author} {\bibfnamefont {L.}~\bibnamefont {Frunzio}}, \bibinfo {author} {\bibfnamefont {S.}~\bibnamefont {Girvin}}, \emph {et~al.},\ }\href@noop {} {\bibfield  {journal} {\bibinfo  {journal} {Nature}\ }\textbf {\bibinfo {volume} {460}},\ \bibinfo {pages} {240} (\bibinfo {year} {2009})}\BibitemShut {NoStop}%
\bibitem [{\citenamefont {Walther}\ \emph {et~al.}(2005)\citenamefont {Walther}, \citenamefont {Resch}, \citenamefont {Rudolph}, \citenamefont {Schenck}, \citenamefont {Weinfurter}, \citenamefont {Vedral}, \citenamefont {Aspelmeyer},\ and\ \citenamefont {Zeilinger}}]{WalResRud2005}%
  \BibitemOpen
  \bibfield  {author} {\bibinfo {author} {\bibfnamefont {P.}~\bibnamefont {Walther}}, \bibinfo {author} {\bibfnamefont {K.~J.}\ \bibnamefont {Resch}}, \bibinfo {author} {\bibfnamefont {T.}~\bibnamefont {Rudolph}}, \bibinfo {author} {\bibfnamefont {E.}~\bibnamefont {Schenck}}, \bibinfo {author} {\bibfnamefont {H.}~\bibnamefont {Weinfurter}}, \bibinfo {author} {\bibfnamefont {V.}~\bibnamefont {Vedral}}, \bibinfo {author} {\bibfnamefont {M.}~\bibnamefont {Aspelmeyer}},\ and\ \bibinfo {author} {\bibfnamefont {A.}~\bibnamefont {Zeilinger}},\ }\href@noop {} {\bibfield  {journal} {\bibinfo  {journal} {Nature}\ }\textbf {\bibinfo {volume} {434}},\ \bibinfo {pages} {169} (\bibinfo {year} {2005})}\BibitemShut {NoStop}%
\bibitem [{\citenamefont {Wang}\ \emph {et~al.}(2021)\citenamefont {Wang}, \citenamefont {Chi}, \citenamefont {Yu}, \citenamefont {Sun}, \citenamefont {Su},\ and\ \citenamefont {Yang}}]{WanChiYu2021}%
  \BibitemOpen
  \bibfield  {author} {\bibinfo {author} {\bibfnamefont {S.-C.}\ \bibnamefont {Wang}}, \bibinfo {author} {\bibfnamefont {Y.}~\bibnamefont {Chi}}, \bibinfo {author} {\bibfnamefont {L.}~\bibnamefont {Yu}}, \bibinfo {author} {\bibfnamefont {Z.}~\bibnamefont {Sun}}, \bibinfo {author} {\bibfnamefont {Q.-P.}\ \bibnamefont {Su}},\ and\ \bibinfo {author} {\bibfnamefont {C.-P.}\ \bibnamefont {Yang}},\ }\href {https://doi.org/10.1103/PhysRevA.103.032413} {\bibfield  {journal} {\bibinfo  {journal} {Phys. Rev. A}\ }\textbf {\bibinfo {volume} {103}},\ \bibinfo {pages} {032413} (\bibinfo {year} {2021})}\BibitemShut {NoStop}%
\bibitem [{\citenamefont {He}\ \emph {et~al.}(2023)\citenamefont {He}, \citenamefont {Zhao}, \citenamefont {Lv}, \citenamefont {Peng}, \citenamefont {Sun}, \citenamefont {Sun}, \citenamefont {Su},\ and\ \citenamefont {Yang}}]{HeZhaLv2023}%
  \BibitemOpen
  \bibfield  {author} {\bibinfo {author} {\bibfnamefont {X.}~\bibnamefont {He}}, \bibinfo {author} {\bibfnamefont {W.-T.}\ \bibnamefont {Zhao}}, \bibinfo {author} {\bibfnamefont {W.-C.}\ \bibnamefont {Lv}}, \bibinfo {author} {\bibfnamefont {C.-H.}\ \bibnamefont {Peng}}, \bibinfo {author} {\bibfnamefont {Z.}~\bibnamefont {Sun}}, \bibinfo {author} {\bibfnamefont {Y.-N.}\ \bibnamefont {Sun}}, \bibinfo {author} {\bibfnamefont {Q.-P.}\ \bibnamefont {Su}},\ and\ \bibinfo {author} {\bibfnamefont {C.-P.}\ \bibnamefont {Yang}},\ }\href {https://doi.org/10.1364/OL.497599} {\bibfield  {journal} {\bibinfo  {journal} {Opt. Lett.}\ }\textbf {\bibinfo {volume} {48}},\ \bibinfo {pages} {4428} (\bibinfo {year} {2023})}\BibitemShut {NoStop}%
\bibitem [{\citenamefont {Bhattacharya}\ \emph {et~al.}(2002)\citenamefont {Bhattacharya}, \citenamefont {van Linden van~den Heuvell},\ and\ \citenamefont {Spreeuw}}]{BhaHeuSpr2002}%
  \BibitemOpen
  \bibfield  {author} {\bibinfo {author} {\bibfnamefont {N.}~\bibnamefont {Bhattacharya}}, \bibinfo {author} {\bibfnamefont {H.~B.}\ \bibnamefont {van Linden van~den Heuvell}},\ and\ \bibinfo {author} {\bibfnamefont {R.~J.~C.}\ \bibnamefont {Spreeuw}},\ }\href {https://doi.org/10.1103/PhysRevLett.88.137901} {\bibfield  {journal} {\bibinfo  {journal} {Phys. Rev. Lett.}\ }\textbf {\bibinfo {volume} {88}},\ \bibinfo {pages} {137901} (\bibinfo {year} {2002})}\BibitemShut {NoStop}%
\bibitem [{\citenamefont {Oka}\ and\ \citenamefont {Kitamura}(2019)}]{OkaSot2019}%
  \BibitemOpen
  \bibfield  {author} {\bibinfo {author} {\bibfnamefont {T.}~\bibnamefont {Oka}}\ and\ \bibinfo {author} {\bibfnamefont {S.}~\bibnamefont {Kitamura}},\ }\href {https://doi.org/https://doi.org/10.1146/annurev-conmatphys-031218-013423} {\bibfield  {journal} {\bibinfo  {journal} {Annual Review of Condensed Matter Physics}\ }\textbf {\bibinfo {volume} {10}},\ \bibinfo {pages} {387} (\bibinfo {year} {2019})}\BibitemShut {NoStop}%
\bibitem [{\citenamefont {Bukov}\ \emph {et~al.}(2015)\citenamefont {Bukov}, \citenamefont {D’Alessio},\ and\ \citenamefont {Polkovnikov}}]{Bukov_2015}%
  \BibitemOpen
  \bibfield  {author} {\bibinfo {author} {\bibfnamefont {M.}~\bibnamefont {Bukov}}, \bibinfo {author} {\bibfnamefont {L.}~\bibnamefont {D’Alessio}},\ and\ \bibinfo {author} {\bibfnamefont {A.}~\bibnamefont {Polkovnikov}},\ }\href {https://doi.org/10.1080/00018732.2015.1055918} {\bibfield  {journal} {\bibinfo  {journal} {Advances in Physics}\ }\textbf {\bibinfo {volume} {64}},\ \bibinfo {pages} {139–226} (\bibinfo {year} {2015})}\BibitemShut {NoStop}%
\bibitem [{\citenamefont {Tiwari}\ \emph {et~al.}(2024)\citenamefont {Tiwari}, \citenamefont {Bhakuni},\ and\ \citenamefont {Sharma}}]{Tiwari_2024}%
  \BibitemOpen
  \bibfield  {author} {\bibinfo {author} {\bibfnamefont {V.}~\bibnamefont {Tiwari}}, \bibinfo {author} {\bibfnamefont {D.~S.}\ \bibnamefont {Bhakuni}},\ and\ \bibinfo {author} {\bibfnamefont {A.}~\bibnamefont {Sharma}},\ }\bibfield  {journal} {\bibinfo  {journal} {Physical Review B}\ }\textbf {\bibinfo {volume} {109}},\ \href {https://doi.org/10.1103/physrevb.109.l161104} {10.1103/physrevb.109.l161104} (\bibinfo {year} {2024})\BibitemShut {NoStop}%
\bibitem [{\citenamefont {Casati}\ and\ \citenamefont {Chirikov}(1989)}]{CasChi-qcbook}%
  \BibitemOpen
  \bibfield  {author} {\bibinfo {author} {\bibfnamefont {G.}~\bibnamefont {Casati}}\ and\ \bibinfo {author} {\bibfnamefont {B.~V.}\ \bibnamefont {Chirikov}},\ }in\ \href@noop {} {\emph {\bibinfo {booktitle} {Quantum Chaos : Between Order and Chaos}}},\ \bibinfo {editor} {edited by\ \bibinfo {editor} {\bibfnamefont {G.}~\bibnamefont {Casati}}\ and\ \bibinfo {editor} {\bibfnamefont {B.~V.}\ \bibnamefont {Chirikov}}}\ (\bibinfo  {publisher} {North-Holland},\ \bibinfo {address} {Amsterdam},\ \bibinfo {year} {1989})\ p.~\bibinfo {pages} {3}\BibitemShut {NoStop}%
\bibitem [{\citenamefont {Collura}\ \emph {et~al.}(2022)\citenamefont {Collura}, \citenamefont {De~Luca}, \citenamefont {Rossini},\ and\ \citenamefont {Lerose}}]{PhysRevX.12.031037}%
  \BibitemOpen
  \bibfield  {author} {\bibinfo {author} {\bibfnamefont {M.}~\bibnamefont {Collura}}, \bibinfo {author} {\bibfnamefont {A.}~\bibnamefont {De~Luca}}, \bibinfo {author} {\bibfnamefont {D.}~\bibnamefont {Rossini}},\ and\ \bibinfo {author} {\bibfnamefont {A.}~\bibnamefont {Lerose}},\ }\href {https://doi.org/10.1103/PhysRevX.12.031037} {\bibfield  {journal} {\bibinfo  {journal} {Phys. Rev. X}\ }\textbf {\bibinfo {volume} {12}},\ \bibinfo {pages} {031037} (\bibinfo {year} {2022})}\BibitemShut {NoStop}%
\bibitem [{\citenamefont {Santhanam}\ \emph {et~al.}(2022)\citenamefont {Santhanam}, \citenamefont {Paul},\ and\ \citenamefont {Kannan}}]{SANTHANAM20221}%
  \BibitemOpen
  \bibfield  {author} {\bibinfo {author} {\bibfnamefont {M.}~\bibnamefont {Santhanam}}, \bibinfo {author} {\bibfnamefont {S.}~\bibnamefont {Paul}},\ and\ \bibinfo {author} {\bibfnamefont {J.~B.}\ \bibnamefont {Kannan}},\ }\href {https://doi.org/https://doi.org/10.1016/j.physrep.2022.01.002} {\bibfield  {journal} {\bibinfo  {journal} {Physics Reports}\ }\textbf {\bibinfo {volume} {956}},\ \bibinfo {pages} {1} (\bibinfo {year} {2022})},\ \bibinfo {note} {quantum kicked rotor and its variants: Chaos, localization and beyond}\BibitemShut {NoStop}%
\bibitem [{\citenamefont {Izrailev}(1990)}]{Izrailev-1990}%
  \BibitemOpen
  \bibfield  {author} {\bibinfo {author} {\bibfnamefont {F.~M.}\ \bibnamefont {Izrailev}},\ }\href {https://doi.org/https://doi.org/10.1016/0370-1573(90)90067-C} {\bibfield  {journal} {\bibinfo  {journal} {Physics Reports}\ }\textbf {\bibinfo {volume} {196}},\ \bibinfo {pages} {299 } (\bibinfo {year} {1990})}\BibitemShut {NoStop}%
\bibitem [{\citenamefont {Chirikov}(1989)}]{Chiri-LesHou}%
  \BibitemOpen
  \bibfield  {author} {\bibinfo {author} {\bibfnamefont {B.~V.}\ \bibnamefont {Chirikov}},\ }in\ \href@noop {} {\emph {\bibinfo {booktitle} {Chaos and Quantum Physics, Les Houches Session LII}}},\ \bibinfo {editor} {edited by\ \bibinfo {editor} {\bibfnamefont {M.-J.}\ \bibnamefont {Giannoni}}, \bibinfo {editor} {\bibfnamefont {A.}~\bibnamefont {Voros}},\ and\ \bibinfo {editor} {\bibfnamefont {J.}~\bibnamefont {Zinn-Justin}}}\ (\bibinfo  {publisher} {North-Holland},\ \bibinfo {address} {Amsterdam},\ \bibinfo {year} {1989})\ p.\ \bibinfo {pages} {443}\BibitemShut {NoStop}%
\bibitem [{\citenamefont {Moore}\ \emph {et~al.}(1994)\citenamefont {Moore}, \citenamefont {Robinson}, \citenamefont {Bharucha}, \citenamefont {Williams},\ and\ \citenamefont {Raizen}}]{MooRobBha1994}%
  \BibitemOpen
  \bibfield  {author} {\bibinfo {author} {\bibfnamefont {F.~L.}\ \bibnamefont {Moore}}, \bibinfo {author} {\bibfnamefont {J.~C.}\ \bibnamefont {Robinson}}, \bibinfo {author} {\bibfnamefont {C.}~\bibnamefont {Bharucha}}, \bibinfo {author} {\bibfnamefont {P.~E.}\ \bibnamefont {Williams}},\ and\ \bibinfo {author} {\bibfnamefont {M.~G.}\ \bibnamefont {Raizen}},\ }\href {https://doi.org/10.1103/PhysRevLett.73.2974} {\bibfield  {journal} {\bibinfo  {journal} {Phys. Rev. Lett.}\ }\textbf {\bibinfo {volume} {73}},\ \bibinfo {pages} {2974} (\bibinfo {year} {1994})}\BibitemShut {NoStop}%
\bibitem [{\citenamefont {Moore}\ \emph {et~al.}(1995)\citenamefont {Moore}, \citenamefont {Robinson}, \citenamefont {Bharucha}, \citenamefont {Sundaram},\ and\ \citenamefont {Raizen}}]{MooRobBha1995}%
  \BibitemOpen
  \bibfield  {author} {\bibinfo {author} {\bibfnamefont {F.~L.}\ \bibnamefont {Moore}}, \bibinfo {author} {\bibfnamefont {J.~C.}\ \bibnamefont {Robinson}}, \bibinfo {author} {\bibfnamefont {C.~F.}\ \bibnamefont {Bharucha}}, \bibinfo {author} {\bibfnamefont {B.}~\bibnamefont {Sundaram}},\ and\ \bibinfo {author} {\bibfnamefont {M.~G.}\ \bibnamefont {Raizen}},\ }\href {https://doi.org/10.1103/PhysRevLett.75.4598} {\bibfield  {journal} {\bibinfo  {journal} {Phys. Rev. Lett.}\ }\textbf {\bibinfo {volume} {75}},\ \bibinfo {pages} {4598} (\bibinfo {year} {1995})}\BibitemShut {NoStop}%
\bibitem [{\citenamefont {Sarkar}\ \emph {et~al.}(2017)\citenamefont {Sarkar}, \citenamefont {Paul}, \citenamefont {Vishwakarma}, \citenamefont {Kumar}, \citenamefont {Verma}, \citenamefont {Sainath}, \citenamefont {Rapol},\ and\ \citenamefont {Santhanam}}]{SarPauVis2017}%
  \BibitemOpen
  \bibfield  {author} {\bibinfo {author} {\bibfnamefont {S.}~\bibnamefont {Sarkar}}, \bibinfo {author} {\bibfnamefont {S.}~\bibnamefont {Paul}}, \bibinfo {author} {\bibfnamefont {C.}~\bibnamefont {Vishwakarma}}, \bibinfo {author} {\bibfnamefont {S.}~\bibnamefont {Kumar}}, \bibinfo {author} {\bibfnamefont {G.}~\bibnamefont {Verma}}, \bibinfo {author} {\bibfnamefont {M.}~\bibnamefont {Sainath}}, \bibinfo {author} {\bibfnamefont {U.~D.}\ \bibnamefont {Rapol}},\ and\ \bibinfo {author} {\bibfnamefont {M.~S.}\ \bibnamefont {Santhanam}},\ }\href {https://doi.org/10.1103/PhysRevLett.118.174101} {\bibfield  {journal} {\bibinfo  {journal} {Phys. Rev. Lett.}\ }\textbf {\bibinfo {volume} {118}},\ \bibinfo {pages} {174101} (\bibinfo {year} {2017})}\BibitemShut {NoStop}%
\bibitem [{\citenamefont {Dadras}\ \emph {et~al.}(2018)\citenamefont {Dadras}, \citenamefont {Gresch}, \citenamefont {Groiseau}, \citenamefont {Wimberger},\ and\ \citenamefont {Summy}}]{Dadras_2018}%
  \BibitemOpen
  \bibfield  {author} {\bibinfo {author} {\bibfnamefont {S.}~\bibnamefont {Dadras}}, \bibinfo {author} {\bibfnamefont {A.}~\bibnamefont {Gresch}}, \bibinfo {author} {\bibfnamefont {C.}~\bibnamefont {Groiseau}}, \bibinfo {author} {\bibfnamefont {S.}~\bibnamefont {Wimberger}},\ and\ \bibinfo {author} {\bibfnamefont {G.~S.}\ \bibnamefont {Summy}},\ }\bibfield  {journal} {\bibinfo  {journal} {Physical Review Letters}\ }\textbf {\bibinfo {volume} {121}},\ \href {https://doi.org/10.1103/physrevlett.121.070402} {10.1103/physrevlett.121.070402} (\bibinfo {year} {2018})\BibitemShut {NoStop}%
\bibitem [{\citenamefont {Dadras}\ \emph {et~al.}(2019{\natexlab{a}})\citenamefont {Dadras}, \citenamefont {Gresch}, \citenamefont {Groiseau}, \citenamefont {Wimberger},\ and\ \citenamefont {Summy}}]{DadGreGro2019}%
  \BibitemOpen
  \bibfield  {author} {\bibinfo {author} {\bibfnamefont {S.}~\bibnamefont {Dadras}}, \bibinfo {author} {\bibfnamefont {A.}~\bibnamefont {Gresch}}, \bibinfo {author} {\bibfnamefont {C.}~\bibnamefont {Groiseau}}, \bibinfo {author} {\bibfnamefont {S.}~\bibnamefont {Wimberger}},\ and\ \bibinfo {author} {\bibfnamefont {G.~S.}\ \bibnamefont {Summy}},\ }\href {https://doi.org/10.1103/PhysRevA.99.043617} {\bibfield  {journal} {\bibinfo  {journal} {Phys. Rev. A}\ }\textbf {\bibinfo {volume} {99}},\ \bibinfo {pages} {043617} (\bibinfo {year} {2019}{\natexlab{a}})}\BibitemShut {NoStop}%
\bibitem [{\citenamefont {Delvecchio}\ \emph {et~al.}(2020{\natexlab{a}})\citenamefont {Delvecchio}, \citenamefont {Petiziol},\ and\ \citenamefont {Wimberger}}]{DelPetWim2020}%
  \BibitemOpen
  \bibfield  {author} {\bibinfo {author} {\bibfnamefont {M.}~\bibnamefont {Delvecchio}}, \bibinfo {author} {\bibfnamefont {F.}~\bibnamefont {Petiziol}},\ and\ \bibinfo {author} {\bibfnamefont {S.}~\bibnamefont {Wimberger}},\ }\bibfield  {journal} {\bibinfo  {journal} {Condensed Matter}\ }\textbf {\bibinfo {volume} {5}},\ \href {https://doi.org/10.3390/condmat5010004} {10.3390/condmat5010004} (\bibinfo {year} {2020}{\natexlab{a}})\BibitemShut {NoStop}%
\bibitem [{\citenamefont {Delvecchio}\ \emph {et~al.}(2020{\natexlab{b}})\citenamefont {Delvecchio}, \citenamefont {Groiseau}, \citenamefont {Petiziol}, \citenamefont {Summy},\ and\ \citenamefont {Wimberger}}]{DelGroPet2020}%
  \BibitemOpen
  \bibfield  {author} {\bibinfo {author} {\bibfnamefont {M.}~\bibnamefont {Delvecchio}}, \bibinfo {author} {\bibfnamefont {C.}~\bibnamefont {Groiseau}}, \bibinfo {author} {\bibfnamefont {F.}~\bibnamefont {Petiziol}}, \bibinfo {author} {\bibfnamefont {G.~S.}\ \bibnamefont {Summy}},\ and\ \bibinfo {author} {\bibfnamefont {S.}~\bibnamefont {Wimberger}},\ }\href {https://doi.org/10.1088/1361-6455/ab63ad} {\bibfield  {journal} {\bibinfo  {journal} {Journal of Physics B: Atomic, Molecular and Optical Physics}\ }\textbf {\bibinfo {volume} {53}},\ \bibinfo {pages} {065301} (\bibinfo {year} {2020}{\natexlab{b}})}\BibitemShut {NoStop}%
\bibitem [{\citenamefont {Boyer}\ \emph {et~al.}(1998)\citenamefont {Boyer}, \citenamefont {Brassard}, \citenamefont {Høyer},\ and\ \citenamefont {Tapp}}]{Boyer_1998}%
  \BibitemOpen
  \bibfield  {author} {\bibinfo {author} {\bibfnamefont {M.}~\bibnamefont {Boyer}}, \bibinfo {author} {\bibfnamefont {G.}~\bibnamefont {Brassard}}, \bibinfo {author} {\bibfnamefont {P.}~\bibnamefont {Høyer}},\ and\ \bibinfo {author} {\bibfnamefont {A.}~\bibnamefont {Tapp}},\ }\href {https://doi.org/10.1002/(sici)1521-3978(199806)46:4/5<493::aid-prop493>3.0.co;2-p} {\bibfield  {journal} {\bibinfo  {journal} {Fortschritte der Physik}\ }\textbf {\bibinfo {volume} {46}},\ \bibinfo {pages} {493–505} (\bibinfo {year} {1998})}\BibitemShut {NoStop}%
\bibitem [{\citenamefont {Fishman}\ \emph {et~al.}(1982)\citenamefont {Fishman}, \citenamefont {Grempel},\ and\ \citenamefont {Prange}}]{PhysRevLett.49.509}%
  \BibitemOpen
  \bibfield  {author} {\bibinfo {author} {\bibfnamefont {S.}~\bibnamefont {Fishman}}, \bibinfo {author} {\bibfnamefont {D.~R.}\ \bibnamefont {Grempel}},\ and\ \bibinfo {author} {\bibfnamefont {R.~E.}\ \bibnamefont {Prange}},\ }\href {https://doi.org/10.1103/PhysRevLett.49.509} {\bibfield  {journal} {\bibinfo  {journal} {Phys. Rev. Lett.}\ }\textbf {\bibinfo {volume} {49}},\ \bibinfo {pages} {509} (\bibinfo {year} {1982})}\BibitemShut {NoStop}%
\bibitem [{\citenamefont {Martin}\ \emph {et~al.}(2008)\citenamefont {Martin}, \citenamefont {Georgeot},\ and\ \citenamefont {Shepelyansky}}]{PhysRevLett.100.044106}%
  \BibitemOpen
  \bibfield  {author} {\bibinfo {author} {\bibfnamefont {J.}~\bibnamefont {Martin}}, \bibinfo {author} {\bibfnamefont {B.}~\bibnamefont {Georgeot}},\ and\ \bibinfo {author} {\bibfnamefont {D.~L.}\ \bibnamefont {Shepelyansky}},\ }\href {https://doi.org/10.1103/PhysRevLett.100.044106} {\bibfield  {journal} {\bibinfo  {journal} {Phys. Rev. Lett.}\ }\textbf {\bibinfo {volume} {100}},\ \bibinfo {pages} {044106} (\bibinfo {year} {2008})}\BibitemShut {NoStop}%
\bibitem [{\citenamefont {Daszuta}\ and\ \citenamefont {Andersen}(2012)}]{PhysRevA.86.043604}%
  \BibitemOpen
  \bibfield  {author} {\bibinfo {author} {\bibfnamefont {B.}~\bibnamefont {Daszuta}}\ and\ \bibinfo {author} {\bibfnamefont {M.~F.}\ \bibnamefont {Andersen}},\ }\href {https://doi.org/10.1103/PhysRevA.86.043604} {\bibfield  {journal} {\bibinfo  {journal} {Phys. Rev. A}\ }\textbf {\bibinfo {volume} {86}},\ \bibinfo {pages} {043604} (\bibinfo {year} {2012})}\BibitemShut {NoStop}%
\bibitem [{\citenamefont {McDowall}\ \emph {et~al.}(2009)\citenamefont {McDowall}, \citenamefont {Hilliard}, \citenamefont {McGovern}, \citenamefont {Grunzweig},\ and\ \citenamefont {Andersen}}]{McDowall2009AFT}%
  \BibitemOpen
  \bibfield  {author} {\bibinfo {author} {\bibfnamefont {P.~D.}\ \bibnamefont {McDowall}}, \bibinfo {author} {\bibfnamefont {A.~J.}\ \bibnamefont {Hilliard}}, \bibinfo {author} {\bibfnamefont {M.}~\bibnamefont {McGovern}}, \bibinfo {author} {\bibfnamefont {T.}~\bibnamefont {Grunzweig}},\ and\ \bibinfo {author} {\bibfnamefont {M.~F.}\ \bibnamefont {Andersen}},\ }\href {https://api.semanticscholar.org/CorpusID:121228661} {\bibfield  {journal} {\bibinfo  {journal} {New Journal of Physics}\ }\textbf {\bibinfo {volume} {11}},\ \bibinfo {pages} {123021} (\bibinfo {year} {2009})}\BibitemShut {NoStop}%
\bibitem [{\citenamefont {Kitaev}(1995)}]{kitaev1995quantum}%
  \BibitemOpen
  \bibfield  {author} {\bibinfo {author} {\bibfnamefont {A.~Y.}\ \bibnamefont {Kitaev}},\ }\href@noop {} {\bibinfo {title} {Quantum measurements and the abelian stabilizer problem}} (\bibinfo {year} {1995}),\ \Eprint {https://arxiv.org/abs/quant-ph/9511026} {arXiv:quant-ph/9511026 [quant-ph]} \BibitemShut {NoStop}%
\bibitem [{\citenamefont {M.A.Nielsen}(2001)}]{Nielsenchuang}%
  \BibitemOpen
  \bibfield  {author} {\bibinfo {author} {\bibfnamefont {I.~C.}\ \bibnamefont {M.A.Nielsen}},\ }\href@noop {} {\bibinfo {title} {Quantum computation and quantum information}} (\bibinfo {year} {2001}),\ \Eprint {https://arxiv.org/abs/978-0521635035} {Cambridge University Press:978-0521635035} \BibitemShut {NoStop}%
\bibitem [{\citenamefont {Ringot}\ \emph {et~al.}(2001)\citenamefont {Ringot}, \citenamefont {Szriftgiser},\ and\ \citenamefont {Garreau}}]{PhysRevA.65.013403}%
  \BibitemOpen
  \bibfield  {author} {\bibinfo {author} {\bibfnamefont {J.}~\bibnamefont {Ringot}}, \bibinfo {author} {\bibfnamefont {P.}~\bibnamefont {Szriftgiser}},\ and\ \bibinfo {author} {\bibfnamefont {J.~C.}\ \bibnamefont {Garreau}},\ }\href {https://doi.org/10.1103/PhysRevA.65.013403} {\bibfield  {journal} {\bibinfo  {journal} {Phys. Rev. A}\ }\textbf {\bibinfo {volume} {65}},\ \bibinfo {pages} {013403} (\bibinfo {year} {2001})}\BibitemShut {NoStop}%
\bibitem [{\citenamefont {Dadras}\ \emph {et~al.}(2019{\natexlab{b}})\citenamefont {Dadras}, \citenamefont {Gresch}, \citenamefont {Groiseau}, \citenamefont {Wimberger},\ and\ \citenamefont {Summy}}]{PhysRevA.99.043617}%
  \BibitemOpen
  \bibfield  {author} {\bibinfo {author} {\bibfnamefont {S.}~\bibnamefont {Dadras}}, \bibinfo {author} {\bibfnamefont {A.}~\bibnamefont {Gresch}}, \bibinfo {author} {\bibfnamefont {C.}~\bibnamefont {Groiseau}}, \bibinfo {author} {\bibfnamefont {S.}~\bibnamefont {Wimberger}},\ and\ \bibinfo {author} {\bibfnamefont {G.~S.}\ \bibnamefont {Summy}},\ }\href {https://doi.org/10.1103/PhysRevA.99.043617} {\bibfield  {journal} {\bibinfo  {journal} {Phys. Rev. A}\ }\textbf {\bibinfo {volume} {99}},\ \bibinfo {pages} {043617} (\bibinfo {year} {2019}{\natexlab{b}})}\BibitemShut {NoStop}%
\bibitem [{\citenamefont {Gu}\ and\ \citenamefont {Franco}(2019)}]{10.1063/1.5099499}%
  \BibitemOpen
  \bibfield  {author} {\bibinfo {author} {\bibfnamefont {B.}~\bibnamefont {Gu}}\ and\ \bibinfo {author} {\bibfnamefont {I.}~\bibnamefont {Franco}},\ }\href {https://doi.org/10.1063/1.5099499} {\bibfield  {journal} {\bibinfo  {journal} {The Journal of Chemical Physics}\ }\textbf {\bibinfo {volume} {151}},\ \bibinfo {pages} {014109} (\bibinfo {year} {2019})}\BibitemShut {NoStop}%
\bibitem [{\citenamefont {Kiely}(2021)}]{article}%
  \BibitemOpen
  \bibfield  {author} {\bibinfo {author} {\bibfnamefont {A.}~\bibnamefont {Kiely}},\ }\href {https://doi.org/10.1209/0295-5075/134/10001} {\bibfield  {journal} {\bibinfo  {journal} {EPL (Europhysics Letters)}\ }\textbf {\bibinfo {volume} {134}},\ \bibinfo {pages} {10001} (\bibinfo {year} {2021})}\BibitemShut {NoStop}%
\bibitem [{\citenamefont {Qin}\ \emph {et~al.}(2017)\citenamefont {Qin}, \citenamefont {Andreanov}, \citenamefont {HC},\ and\ \citenamefont {Flach}}]{InteratingQKRs}%
  \BibitemOpen
  \bibfield  {author} {\bibinfo {author} {\bibfnamefont {P.}~\bibnamefont {Qin}}, \bibinfo {author} {\bibfnamefont {A.}~\bibnamefont {Andreanov}}, \bibinfo {author} {\bibfnamefont {H.~C.~P.}\ \bibnamefont {HC}},\ and\ \bibinfo {author} {\bibfnamefont {S.}~\bibnamefont {Flach}},\ }\href {https://doi.org/10.1038/srep41139} {\bibinfo {title} {Interacting ultracold atomic kicked rotors: loss of dynamical localization}} (\bibinfo {year} {2017})\BibitemShut {NoStop}%
\end{thebibliography}%
\end{document}